\pdfoutput=1

\documentclass[11pt]{article}

\usepackage{ACL2023}

\usepackage{times}
\usepackage{latexsym}

\usepackage[T1]{fontenc}

\usepackage[utf8]{inputenc}

\usepackage{microtype}

\usepackage{inconsolata}

\usepackage{algorithm}
\usepackage{algpseudocode}
\usepackage{amsmath}
\usepackage{tabularx}
\usepackage{booktabs}
\usepackage{subcaption}
\usepackage{multirow}
\usepackage{amsfonts,amssymb,bbding,pifont,xcolor,makecell}

\usepackage{newfloat}
\usepackage{listings}
\usepackage{graphicx}
\usepackage{xurl}
\usepackage{xparse}
\usepackage{array}
\usepackage{stfloats}

\usepackage{helvet}
\usepackage{courier}
\usepackage{colortbl}  
\usepackage[normalem]{ulem}
\usepackage{ragged2e}

\newcommand{\blue}[1]{\textcolor{blue}{#1}}
\newcommand{\cellhl}{\cellcolor{blue!15}}

\title{Hearing Lips in Noise: Universal Viseme-Phoneme Mapping and Transfer for Robust Audio-Visual Speech Recognition}

\author{Yuchen Hu$^1$, Ruizhe Li$^2$, Chen Chen$^1$, Chengwei Qin$^1$, Qiushi Zhu$^{3}$, Eng Siong Chng$^1$ \\
$^1$Nanyang Technological University, Singapore
\quad$^2$University of Aberdeen, UK \\
$^3$University of Science and Technology of China, China \\
\small{\texttt{\{yuchen005@e., chen1436@e., chengwei003@e., aseschng@\}ntu.edu.sg,}} \\ \small{\texttt{ruizhe.li@abdn.ac.uk, qszhu@mail.ustc.edu.cn}}
}

\begin{document}
\maketitle
\begin{abstract}
Audio-visual speech recognition (AVSR) provides a promising solution to ameliorate the noise-robustness of audio-only speech recognition with visual information.
However, most existing efforts still focus on audio modality to improve robustness considering its dominance in AVSR task, with noise adaptation techniques such as front-end denoise processing.
Though effective, these methods are usually faced with two practical challenges: 1) lack of sufficient labeled noisy audio-visual training data in some real-world scenarios and 2) less optimal model generality to unseen testing noises.
In this work, we investigate the noise-invariant visual modality to strengthen robustness of AVSR, which can adapt to any testing noises while without dependence on noisy training data, \textit{a.k.a.}, unsupervised noise adaptation.
Inspired by human perception mechanism, we propose a universal viseme-phoneme mapping (UniVPM) approach to implement modality transfer, which can restore clean audio from visual signals to enable speech recognition under any noisy conditions.
Extensive experiments on public benchmarks LRS3 and LRS2 show that our approach achieves the state-of-the-art under various noisy as well as clean conditions.
In addition, we also outperform previous state-of-the-arts on visual speech recognition task\footnote{Code is available at \url{https://github.com/YUCHEN005/UniVPM}.}.


\end{abstract}

\section{Introduction}
\label{sec:intro}
The world surrounding us involves multiple modalities, including vision, audio, text, etc., which complement each other and jointly comprise human perception~\citep{baltruvsaitis2018multimodal,zhu2021deep}.
Audio-visual speech recognition (AVSR) leverages both audio and visual modalities to understand human speech, which provides a promising solution to ameliorate the noise-robustness of audio-only speech recognition with noise-invariant lip movement information~\citep{sumby1954visual}.

\begin{figure}[t]
\centering
    \includegraphics[width=1.0\columnwidth]{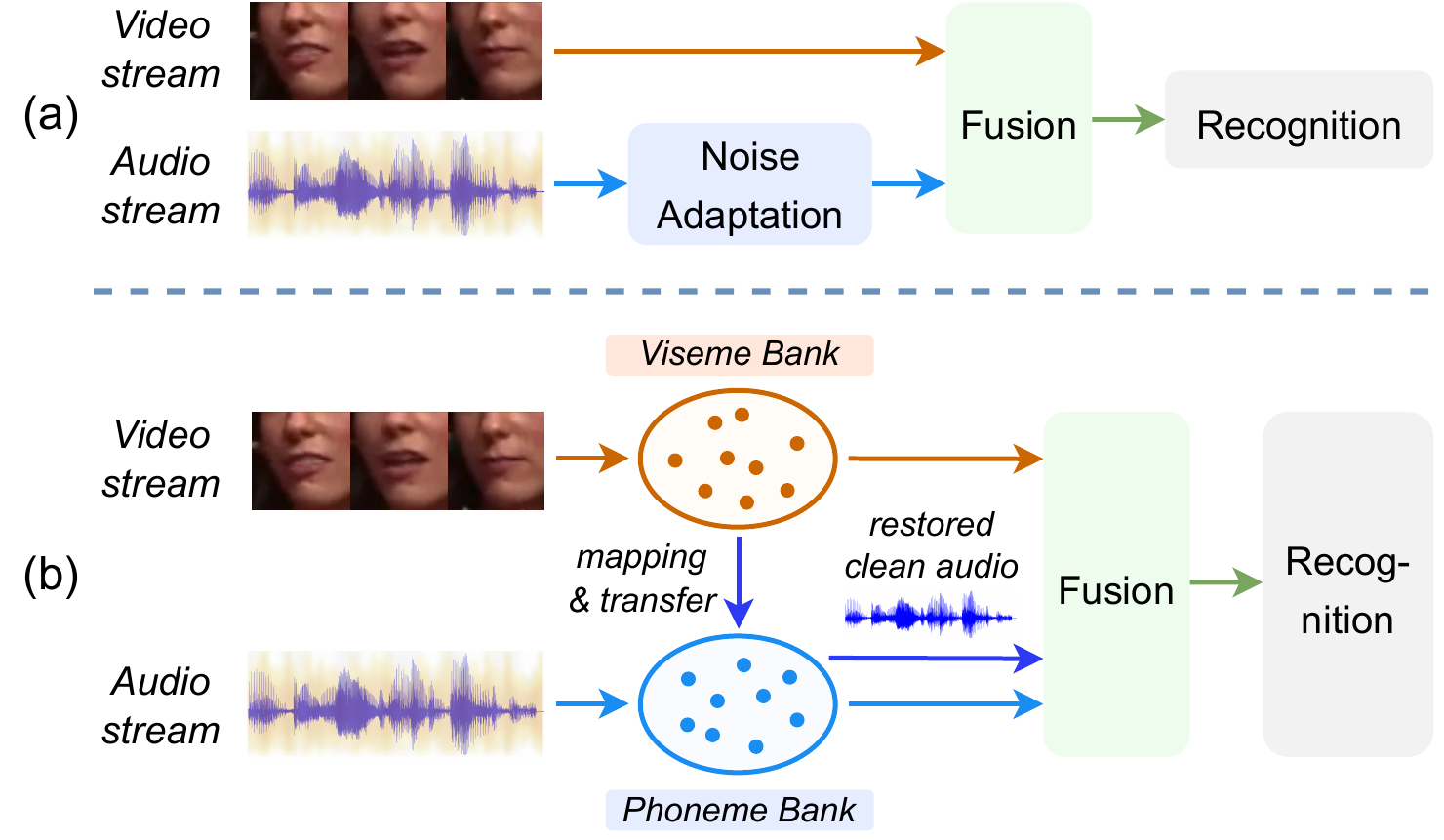}
    \vspace{-0.6cm}
    \caption{Illustration of noisy audio-visual speech recognition. (a) Mainstream AVSR approaches with noise adaptation. (b) Our framework constructs viseme-phoneme mapping for modality transfer, which restores clean audio from visual signals to enable speech recognition under any noisy conditions.}\label{fig1}
    \vspace{-0.4cm}
\end{figure}

However, most existing efforts still focus on audio modality to improve noise-robustness considering its dominance in AVSR, where audio modality contains much richer information to represent speech content than visual modality~\citep{sataloff1992human,ren2021learning}.
Current mainstream approaches introduce noise adaptation techniques to improve robustness\footnote{Experimental analysis are in~\S\ref{assec:analysis_robustness_avsr} and~\S\ref{sssec:unseen_noise}.\label{fn2}}, inspired by robust speech recognition~\citep{wang2020complex}.
Most of them leverage noise-corrupted training data to strengthen robustness~\citep{afouras2018deep,ma2021end,song2022multimodal}, and recent works extend it to self-supervised learning scheme~\citep{shi2022robust,hsu2022u}. Based on that, latest works introduce speech enhancement as front-end to denoise before recognition~\citep{xu2020discriminative,hong2022visual}.
Despite the effectiveness, these methods are usually faced with two practical challenges.
First, they require abundant labeled noisy audio-visual data for network training, which is not always available in some real-world scenarios~\citep{lin2021unsupervised,chen2022noise}.
Second, the well-trained model may not adapt to new-coming noise scenes in practical applications\textsuperscript{\ref{fn2}}, resulting in less optimal model generality~\citep{meng2017unsupervised}.
Therefore, our research idea in this paper is leveraging visual modality to develop a general noise-robust AVSR system while without dependence on noisy training data.

We may gain some inspirations from human perception mechanism of noisy audio-visual speech.
Neuroscience studies~\citep{nath2011dynamic} find that human brain will unconsciously rely more on the lip movement to understand speech under noisy conditions (\textit{a.k.a.}, McGurk Effect,~\citealp{mcgurk1976hearing}). 
During this process, instead of directly recognizing lip movement, human brain will first transfer it to speech signal in auditory cortex for further understanding~\citep{bourguignon2020lip,megevand2020crossmodal}.
With prior knowledge of lip-audio mapping, human brain can restore informative clean audio from lip movement under any noisy conditions to aid in speech understanding~\citep{bernstein2004auditory,aller2022differential}.

Motivated by above observations, we propose a universal viseme-phoneme\footnote{Phoneme is the phonetic base unit (from clean audio), and viseme is the visual equivalent of phoneme.} mapping approach (UniVPM) to implement modality transfer, which can restore clean audio from lip movement to enable speech recognition under any noisy conditions.
We first build two universal memory banks to model all the visemes and phonemes via online balanced clustering.
Based on that, an adversarial mutual information estimator is proposed to construct strong viseme-phoneme mapping, which enables final lip-to-audio modality transfer via retrieval.
As a result, our system can adapt well to any testing noises while without noisy training data.
Empirical results show the effectiveness of our approach.
Our contributions are summarized as:

\begin{itemize}
\item We present UniVPM, a general noise-robust AVSR approach investigated on visual modality, which can adapt to any testing noises while without dependence on noisy training data, \textit{a.k.a.}, unsupervised noise adaptation.

\item We build two universal banks to model all the visemes and phonemes via online balanced clustering, followed by an adversarial mutual information estimator to construct strong mapping between them, which enables modality transfer to restore clean audio from lip movement for speech recognition under any noises.

\item Our UniVPM outperforms previous state-of-the-arts on LRS3 and LRS2 benchmarks. Extensive experiments also show its superiority on visual speech recognition (VSR) task.
\end{itemize}

\section{Related Work}
\label{sec:related_work}
\noindent\textbf{Audio-Visual Speech Recognition.}
AVSR provides a promising solution to noise-robust speech recognition with the noise-invariant visual modality~\citep{afouras2018deep}.
However, most existing efforts still focus on audio modality to improve robustness considering its dominance in AVSR task~\citep{sataloff1992human,ren2021learning}.
Mainstream approaches introduce noise adaptation techniques to strengthen robustness, where most of them leverage noise-corrupted data to improve network training~\citep{afouras2018deep,ma2021end,pan2022leveraging,shi2022robust,hsu2022u}, and recent works further introduce speech enhancement as front-end to denoise before recognition~\citep{xu2020discriminative,hong2022visual}.
Despite the effectiveness, these methods require abundant labeled noisy audio-visual training data that is not always available in some real scenarios, and they may not adapt to the new-coming noise scenes in practical applications.
In this work, we investigate the visual modality to develop a general noise-robust AVSR approach while without dependence on noisy training data, \textit{a.k.a.}, unsupervised noise adaptation.

\label{sec:related_work:memory_network}
\noindent\textbf{Memory Network.}
Memory network~\citep{weston2014memory} presents a long-term memory component that can be read from and written in with inference capability. 
\citet{miller2016key} introduces key-value memory structure where key memory is used to address a query and the retrieved output is obtained from value memory using the address.
Since this scheme can remember selected information, it is effective for augmenting features in many tasks, including video prediction~\citep{lee2021video}, cross-modal retrieval~\citep{song2018deep,chen2020imram}, lip reading~\citep{kim2021multi,kim2022distinguishing} and talking face generation~\citep{park2022synctalkface}.
Despite the advances, the memory network is prone to over-fitting when handling imbalanced distributed data, \textit{a.k.a.}, long tail\footnote{Phoneme distribution in English is a long tail, see \S\ref{assec:analysis_phoneme_distribution}.}~\citep{liu2019large}, which may fail to model the minority classes well.
In this work, we propose to build two memory banks via online balanced clustering to model all the visemes and phonemes equally, \textit{i.e.}, universal.

\noindent\textbf{Viseme-Phoneme Mapping.}
Viseme-phoneme mapping is important to many visual-audio learning tasks, including speech recognition~\citep{chan2022multi}, lip reading~\citep{ren2021learning} and lip-to-speech synthesis~\citep{prajwal2020learning}.
Among them, cross-modal distillation is a popular technique to transfer knowledge from viseme to phoneme~\citep{afouras2020asr,zhao2020hearing,ren2021learning}.
Other works design specific neural networks to learn their mapping~\citep{qu2019lipsound,kim2021lip}.
Recent studies introduce self-supervised learning to capture correlations between visemes and phonemes~\citep{qu2021lipsound2,ma2021lira}.
Though effective, these methods are often challenged by the ambiguity of homophenes~\citep{bear2017phoneme} where one lip shape can produce different sounds.
To this end, we propose an adversarial mutual information estimator to construct strict viseme-phoneme mapping with the strong distinguishing ability of adversarial learning.

\section{Methodology}
\label{sec:method}

\begin{figure*}[t]
\centering
    \includegraphics[width=0.88\textwidth]{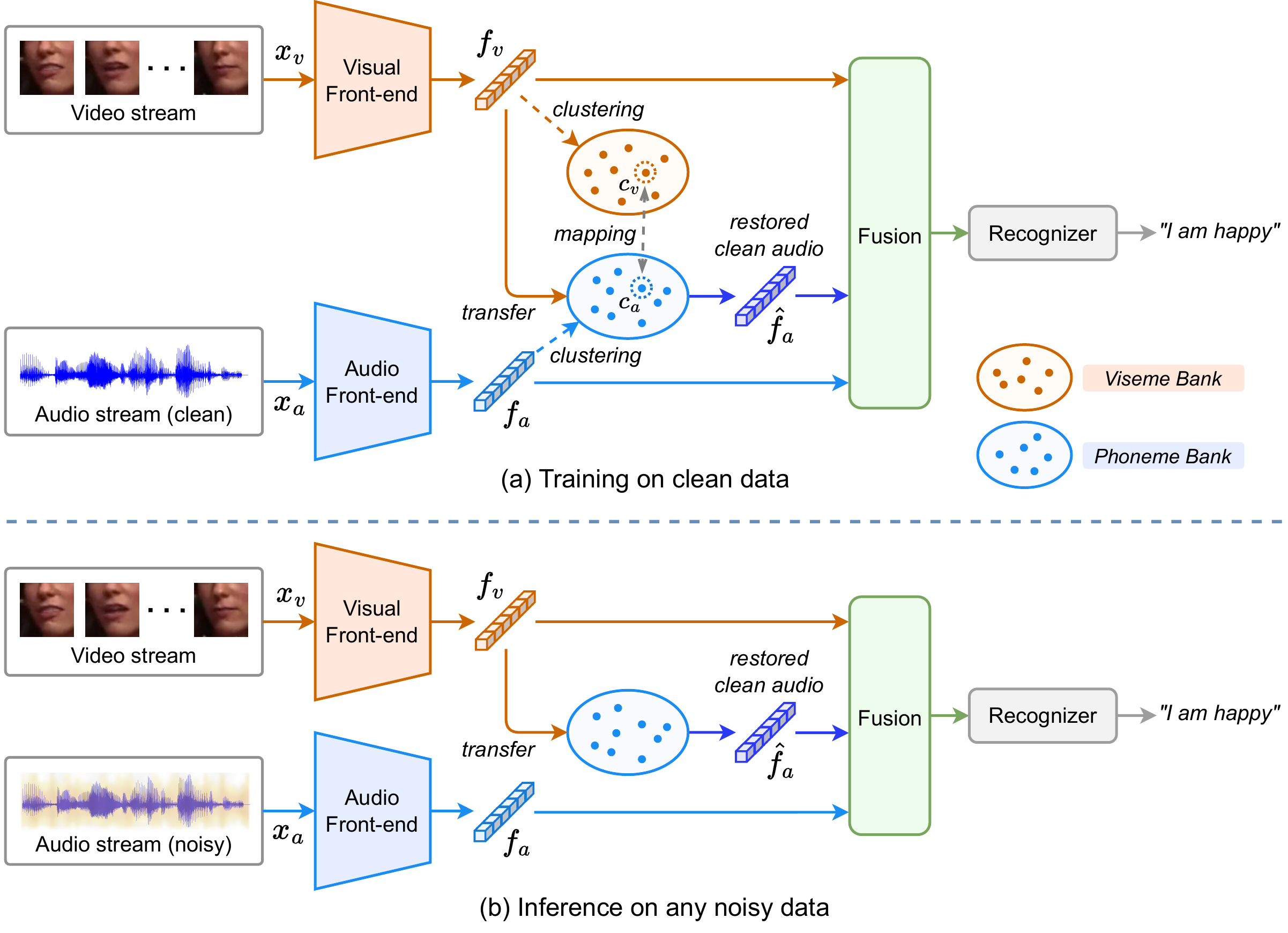}
    \vspace{-0.2cm}
    \caption{Illustration of our proposed UniVPM. (a) Training on clean audio-visual data to construct universal viseme-phoneme mapping. (b) Inference on any noisy data with restored clean audio from modality transfer.}\label{fig2}
    \vspace{-0.3cm}
\end{figure*}

\subsection{Overview}
\label{ssec:overview}
The overall framework of proposed UniVPM is illustrated in Fig.~\ref{fig2}.
During training, we first send the input video and clean audio streams into two front-ends for processing, which generates modality sequences $f_v, f_a \in \mathbb{R}^{T\times D}$, where $T$ is number of frames and $D$ is embedding dimension.
These frames are sent into two memory banks to model all the visemes and phonemes, using an online balanced clustering algorithm where each cluster center represents a specific viseme or phoneme.
Then, we propose an adversarial mutual information estimator to construct strong mapping between corresponding visemes and phonemes.
Based on that, we finally implement modality transfer via retrieval to restore clean audio from visual signals, which enables speech recognition under any testing noises.

\subsection{Online Balanced Clustering}
\label{ssec:ob_clustering}
\begin{figure}[t]
\centering
    \includegraphics[width=1.0\columnwidth]{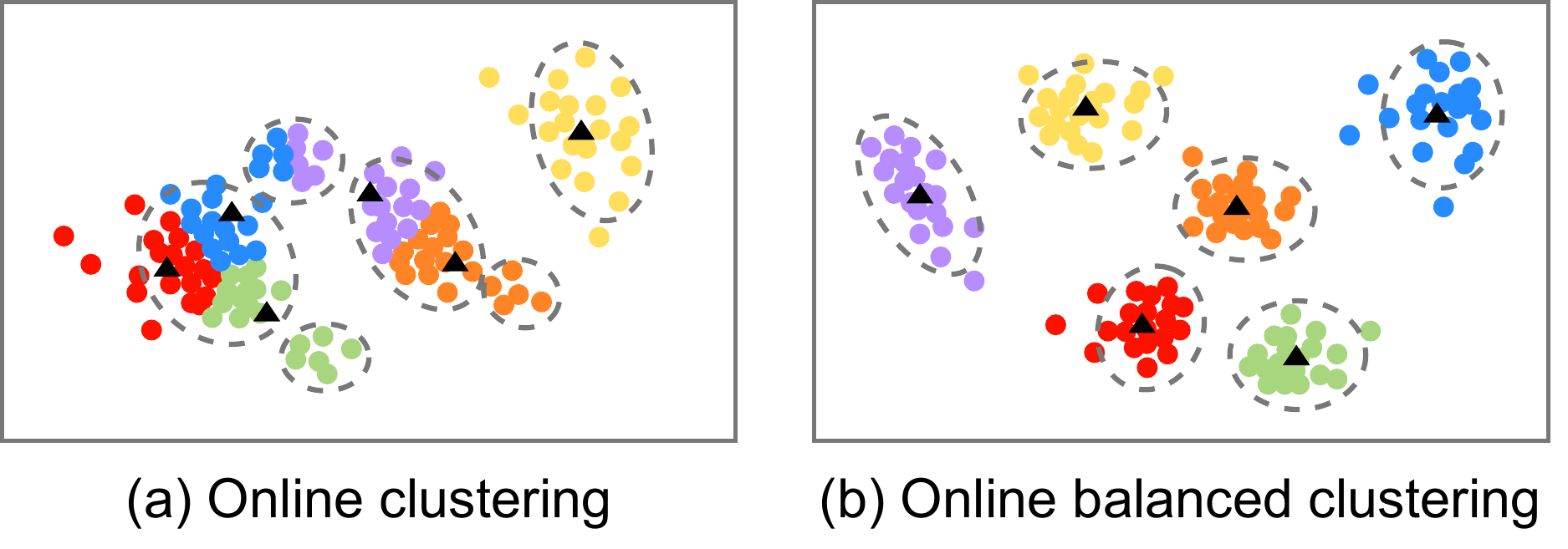}
    \vspace{-0.6cm}
    \caption{t-SNE visualization of clustered phonemes from (a) online clustering (with random pruning to keep fixed cluster size, details are in \S\ref{assec:detail_online_cluster}), and (b) our proposed online balanced clustering.
    We randomly select six clusters for visualization, and black triangle denotes the cluster center.
    Dashed ellipses highlight the real phoneme classes, which are confirmed by pre-trained phoneme recognition model~\citep{phy22phoneme}.}\label{fig3}
    \vspace{-0.3cm}
\end{figure}

Clustering is a widely used knowledge discovery technique to partition a set of data points into homogeneous groups, which has a variety of applications such as data mining~\citep{fayyad1996advances}.
Among them, $K$-Means algorithm~\citep{macqueen1967classification} is the most well-known and popular one.
However, it cannot be directly applied for our viseme and phoneme clustering due to imbalanced data distribution (see \S\ref{assec:analysis_phoneme_distribution}).
This may challenge $K$-Means clustering according to uniform effect~\citep{xiong2006k}.
As shown in Fig.~\ref{fig3} (a), most cluster centers gather in the majority data class (\textit{i.e.}, over-fitting), leaving the minority class not well modeled.

\begin{algorithm}[t]
\caption{\small Online Balanced Clustering.}
\label{alg1}
\small 
\begin{algorithmic}[1]
    \Require Streaming data $D$, number of clusters $N$, maximum cluster size $S_{max}$.
    \State Initialize an empty memory bank $\mathcal{B}$ and a list of empty cluster banks $\{\mathcal{B}_1, \mathcal{B}_2, ..., \mathcal{B}_N\}$.
    \While{$len(\mathcal{B}) \leq N$}
        \State Receive new batch data $d$ from $D$
        \State Append all frame samples in $d$ to bank $\mathcal{B}$
    \EndWhile
    \State Initialize a list of cluster centers $\{c_1, c_2, ..., c_N\}$ from $\mathcal{B}$ using \textsc{K-Means++} Algorithm~\citeyearpar{arthur2006k}
    \For{batch data $d \in D$}
        \State Append all frame samples in $d$ to bank $\mathcal{B}$
        \State $\{\mathcal{B}_1, ..., \mathcal{B}_N\} = \textsc{Re-allocate}(\mathcal{B}, \{c_1, ..., c_N\})$ 
        \State $\{c_1, ..., c_N\} = \textsc{Renew-centers}(\{\mathcal{B}_1, ..., \mathcal{B}_N\})$
        \State Calculate average cluster size $S_{avg} = len(\mathcal{B})/N $
        \State Threshold cluster size $S_{thr} = \min(S_{avg}, S_{max})$
        \For{$i = 1, 2, ..., N$}
            \If{$len(\mathcal{B}_i) > S_{thr}$} \Comment{\blue{\textsc{Undersampling}}}
                \State Maintain the $S_{thr}$-nearest samples to $c_i$ in $\mathcal{B}_i$
                \State Update $\mathcal{B}$ accordingly
            \Else \Comment{\blue{\textsc{Oversampling}}}
                \State Set a random weight $\alpha \in (0, 1)$
                \State Find the nearest sample $d_{near}$ to $c_i$ in $\mathcal{B}_i$
                \State $d_{new} = d_{near} \cdot \alpha + c_i \cdot (1-\alpha)$
                \State $\mathcal{B}_i.append(d_{new})$
                \State Update $\mathcal{B}$ accordingly
            \EndIf
        \EndFor
    \EndFor
\end{algorithmic}
\normalsize
\end{algorithm}

To this end, we propose an Online Balanced Clustering algorithm in Alg.~\ref{alg1} to model all the visemes and phonemes equally from input frames.
First, we set the number of clusters $N$ to $40$, following the amount of English phonemes~\citep{phy22phoneme}.
Then, we set a maximum cluster size $S_{max}$ (\textit{i.e.}, number of samples in each cluster) to control the total memory.
We also initialize an empty bank $\mathcal{B}$ as an overall cache, as well as a list of empty banks $\{\mathcal{B}_1, \mathcal{B}_2, ..., \mathcal{B}_N\}$ to cache each cluster.

The proposed algorithm is executed in three steps, center initialization, $K$-Means clustering and re-sampling.
First, we collect the first few batches of data frames into $\mathcal{B}$ to initialize $N$ dispersed cluster centers $\{c_1, c_2, ..., c_N\}$, using $K$-Means++ algorithm~\citep{arthur2006k}.
Second, we add the current batch data to bank $\mathcal{B}$ and employ vanilla $K$-Means algorithm to re-allocate each sample in the bank to the nearest cluster center, after that the new cluster centers would be updated.
Finally, we propose a re-sampling strategy to balance the size of different clusters as well as control the total memory of bank $\mathcal{B}$, by setting a threshold cluster size $S_{thr}$ (line 12 in Alg.~\ref{alg1}).
For those clusters with more than $S_{thr}$ samples (\textit{i.e.}, majority cluster), we perform undersampling by only maintaining the $S_{thr}$ nearest samples to cluster center.
In contrast, for the minority clusters with less samples than threshold, we propose oversampling to interpolate a new sample between center and the nearest sample with a random weight, inspired by SMOTE algorithm~\citep{chawla2002smote}.
In this way, as illustrated in Fig.~\ref{fig3} (b), the resulted clusters would be balanced-sized and separated to better represent each of the visemes and phonemes.

\subsection{Adversarial Mutual Information Estimator}
\label{ssec:amie}
After clustering visemes and phonemes in banks, we propose an Adversarial Mutual Information Estimator (AMIE) to construct strong mapping between them.
Mutual Information (MI) is a commonly used measure to explore the coherence between two distributions, which is, however, historically difficult to estimate.
Recently,~\citet{belghazi2018mutual} propose a Mutual Information Neural Estimation (MINE) approach to approximate MI lower bound with neural network.
Based on that, we propose an adversarial learning approach to maximize the MI between visemes and phonemes, in order to construct strict mapping between them and thus alleviate the ambiguity of homophenes.

\subsubsection{Preliminary Theory of MINE}
\label{sssec:pre_theory}
Mutual information measures the mutual dependency between two probability distributions,
\begin{equation}
\label{eq1}
\begin{aligned}
    I(X, Y) &= \sum_{x\in X} \sum_{y\in Y} p(x, y)\log\frac{p(x, y)}{p(x)p(y)},
\end{aligned}
\end{equation}
where $p(x, y)$ is the joint probability distribution of $X$ and $Y$, and $p(x)$ and $p(y)$ are the marginals.

Therefore, the mutual information can be written in terms of  Kullback-Leibler (KL-) divergence:
\begin{equation}
\label{eq2}
\begin{aligned}
    I(X, Y) &= D_{K\hspace{-0.02cm}L}(p(x, y) \hspace{0.1cm}\Vert\hspace{0.1cm} p(x)p(y)),  
\end{aligned}
\end{equation}
where $D_{K\hspace{-0.02cm}L}$ is defined as:
\begin{equation}
\label{eq3}
\begin{aligned}
    D_{K\hspace{-0.02cm}L}(p\hspace{0.1cm}\Vert\hspace{0.1cm} q) &= \sum_{x\in X} p(x)\log\frac{p(x)}{q(x)}, 
\end{aligned}
\end{equation}

Furthermore, the $KL$-divergence admits the Donsker-Varadhan (DV) representation~\citep{donsker1983asymptotic,belghazi2018mutual}: 
\begin{equation}
\label{eq4}
\begin{aligned}
    D_{K\hspace{-0.02cm}L}(p\hspace{0.1cm}\Vert\hspace{0.1cm} q) &= \sup_{T:\Omega\rightarrow\mathbb{R}} \mathbb{E}_p[T] - \log(\mathbb{E}_q[e^T]),
\end{aligned}
\end{equation}
where the supremum is taken over all functions $T$ on $\Omega\subset\mathbb{R}^d$ to guarantee two finite expectations.
Therefore, we have the MI lower bound:
\begin{equation}
\label{eq5}
\begin{aligned}
    I(X, Y) &\geq I_\Theta(X, Y), 
\end{aligned}
\end{equation}
where $I_\Theta$ is the neural information measure,
\begin{equation}
\label{eq6}
\begin{aligned}
    I_\Theta(X, Y) &= \sup_{\theta\in \Theta} \mathbb{E}_{p(x, y)}[T_\theta(x, y)] \\
    &- \log(\mathbb{E}_{p(x)p(y)}[e^{T_\theta(x, y)}]), 
\end{aligned}
\end{equation}
and $T_\theta$ denotes a trainable neural network.

\subsubsection{Proposed AMIE}
\label{sssec:proposed_amie}
Based on MINE, we propose an Adversarial Mutual Information Estimator to explore and maximize the mutual information between clustered visemes and phonemes.
As illustrated in Fig.~\ref{fig2} and~\ref{fig4}, given a visual sequence $f_v$, we send each frame of it into viseme bank to find the nearest cluster center $c_v$, which forms the viseme sequence $s_v \in \mathbb{R}^{T\times D}$.
Similarly, we obtain a phoneme sequence $s_a$ to represent audio features $f_a$.
The neural network $T_\theta$ then feeds $\{s_v, s_a\}$ to output a scalar for MI estimation, where $T_\theta$ is a 3-layer classifier with output as a 1-dimensional scalar.
Furthermore, since we do not concern the accurate value of MI when maximizing it, we employ Jensen-Shannon (JS) representation~\citep{hjelm2018learning} to approximate $KL$-divergence in Eq.~\ref{eq4}, which has been proved with more stable neural network optimization.
Therefore, the mutual information between clustered visemes and phonemes is estimated as:
\begin{equation}
\label{eq7}
\begin{aligned}
    I_\Theta^{J\hspace{-0.02cm}S}(s_v, s_a) = \sup_{\theta\in \Theta} &\mathbb{E}_{p(s_v, s_a)}[-\text{sp}(-T_\theta(s_v, s_a))] \\
    - &\mathbb{E}_{p(s_v)p(s_a)}[\text{sp}(T_\theta(s_v, \tilde{s}_a))], 
\end{aligned}
\end{equation}
where $\tilde{s}_a$ is the shuffle-ordered version of $s_a$ that subjects to the marginal distributions of phonemes, and $\text{sp}(z) = \log(1 + e^z)$ is the softplus function.

As stated in~\citet{belghazi2018mutual}, the neural network $T_\theta$ can be used to estimate MI between generated data ($s_v, s_a$ in our case) by directly trained on them.
However, this will suffer a lot from the poor quality of generated data at early training stage.
One feasible scheme~\citep{zhu2021arbitrary} is to train $T_\theta$ on real data ($f_v, f_a$ in our case) and then estimate MI on generated data, but this suffers from the ambiguity of homophenes (see Fig.~\ref{fig8}).
To this end, we propose AMIE with adversarial learning to estimate and maximize the MI between corresponding visemes and phonemes, which can construct strict viseme-phoneme mapping without ambiguity.

Inspired by GAN~\citep{Goodfellow2014}, we design the AMIE as discriminator and the viseme-phoneme banks as generator.
Based on that, the adversarial loss is defined as:
\begin{equation}
\label{eq8}
\begin{aligned}
    \mathcal{L}_{G\hspace{-0.02cm}A\hspace{-0.02cm}N} &= \mathcal{L}_{D} +  \mathcal{L}_{G} \\ 
    &= I_\Theta^{J\hspace{-0.02cm}S}(f_v, f_a) + [-I_\Theta^{J\hspace{-0.02cm}S}(s_v, s_a)],
\end{aligned}
\end{equation}

Our framework employs an adversarial learning strategy for optimization, where $D$ and $G$ play a two-player minimax game as detailed in Alg.~\ref{alg2}.
As a result, the estimated MI between corresponding visemes and phonemes would be maximized to construct mapping relationships.
The strong distinguishing ability of adversarial learning enables strict viseme-phoneme mapping to overcome the ambiguity of homophenes, as shown in Fig.~\ref{fig5}.


\subsection{Modality Transfer}
\label{ssec:modal_transfer}

\begin{figure}[t]
\centering
    \includegraphics[width=0.95\columnwidth]{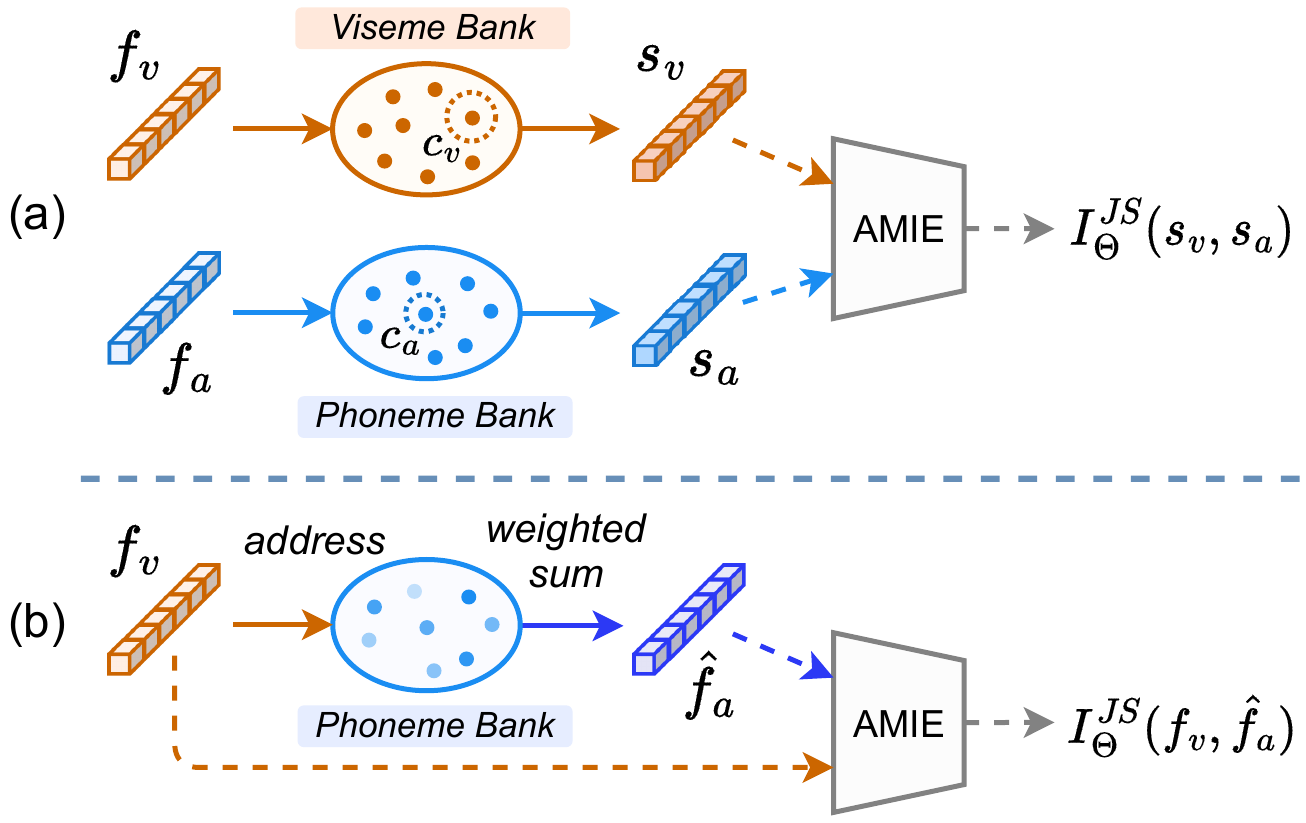}
    \vspace{-0.25cm}
    \caption{Illustration of (a) viseme-phoneme mapping via AMIE, and (b) modality transfer via retrieval.}\label{fig4}
    \vspace{-0.35cm}
\end{figure}

With constructed viseme-phoneme mapping, we can finally implement modality transfer to restore clean audio from lips.
As shown in Fig.~\ref{fig4}, given the visual sequence $f_v$ and clustered phoneme centers $\{c_a^1, c_a^2, ..., c_a^N\}$, we calculate an addressing score $\mathcal{A}^{i,j}$ to indicate the probability that the $i$-th visual frame corresponds to the $j$-th phoneme cluster:
\begin{equation}
\label{eq9}
\begin{aligned}
    \mathcal{A}^{i,j} &= \frac{\exp(\langle f_v^i, c_a^j\rangle/\tau)}{\sum_{k=1}^{N} \exp(\langle f_v^i, c_a^k\rangle/\tau)},
\end{aligned}
\end{equation}
where $\langle\hspace{0.04cm}\cdot, \cdot\hspace{0.04cm}\rangle$ denotes cosine similarity, $\tau$ is temperature weight.
The restored clean audio frames are:
\begin{equation}
\label{eq10}
\begin{aligned}
    \hat{f}_a^i &= \sum_{j=1}^{N}\mathcal{A}^{i,j}\cdot c_a^j,
\end{aligned}
\end{equation}

To supervise the quality of restored audio $\hat{f}_a = \{\hat{f}_a^i\}_{i=1}^{T}$, we first employ AMIE to maximize the MI between $\hat{f}_a$ and $f_v$, where Eq.~\ref{eq8} is rewritten as:
\begin{equation}
\label{eq11}
\small
\begin{aligned}
    \mathcal{L}_{G\hspace{-0.02cm}A\hspace{-0.02cm}N} &= I_\Theta^{J\hspace{-0.02cm}S}(f_v, f_a) + [-I_\Theta^{J\hspace{-0.02cm}S}(s_v, s_a)-I_\Theta^{J\hspace{-0.02cm}S}(f_v, \hat{f}_a)],
\end{aligned}
\normalsize
\end{equation}
along with a reconstruction loss $\mathcal{L}_{rec} = \Vert \hat{f}_a - f_a\Vert_2$ to enable restoration of high-quality clean audio.

\subsection{Optimization}
\label{ssec:optim}
The UniVPM is optimized in an end-to-end manner (see Alg.~\ref{alg2}), with the final training objective as:
\begin{equation}
\label{eq12}
\small
\begin{aligned}
    \mathcal{L} &= \mathcal{L}_{A\hspace{-0.01cm}S\hspace{-0.01cm}R} + \lambda_{G\hspace{-0.02cm}A\hspace{-0.02cm}N} \cdot \mathcal{L}_{G\hspace{-0.02cm}A\hspace{-0.02cm}N} + \lambda_{rec} \cdot \mathcal{L}_{rec} + \lambda_{var} \cdot \mathcal{L}_{var},
\end{aligned}
\normalsize
\end{equation}
where $\mathcal{L}_{A\hspace{-0.01cm}S\hspace{-0.01cm}R}$ denotes the downstream speech recognition loss. 
$\mathcal{L}_{var}$ is a variance regularization term to disperse the clustered viseme and phoneme centers, which aims to ease their mapping construction.
$\lambda_{G\hspace{-0.02cm}A\hspace{-0.02cm}N}$, $\lambda_{rec}$ and $\lambda_{var}$ are weighting parameters.

\begin{table*}[t]
\centering
\resizebox{1.0\textwidth}{!}{
\begin{tabular}{ccc|cccccc|cccccc|cccccc|c}
\toprule
\multirow{2}{*}{Method} & PT & FT & \multicolumn{6}{c|}{Babble, SNR (dB) =} & \multicolumn{6}{c|}{Speech, SNR (dB) =} & \multicolumn{6}{c|}{Music + Natural, SNR (dB) =} & Clean \\ 
& Type & Type & -10 & -5 & 0 & 5 & 10 & avg & -10 & -5 & 0 & 5 & 10 & avg & -10 & -5 & 0 & 5 & 10 & avg & $\infty$  \\ 
\midrule
RNN-T~\citeyearpar{makino2019recurrent} & - & Clean & - & - & - & - & - & - & - & - & - & - & - & - & - & - & - & - & - & - & 4.5 \\
Hyb-AVSR~\citeyearpar{ma2021end} & - & Noisy & - & - & - & - & - & - & - & - & - & - & - & - & - & - & - & - & - & - & 2.3 \\
TM-seq2seq~\citeyearpar{afouras2018deep} & - & Noisy & - & - & 42.5 & - & - & - & - & - & - & - & - & - & - & - & - & - & - & - & 7.2 \\
EG-seq2seq~\citeyearpar{xu2020discriminative} & - & Noisy & 38.6 & 31.1 & 25.5 & 24.3 & 20.7 & 28.0 & - & - & - & - & - & - & - & - & - & - & - & - & 6.8 \\
u-HuBERT~\citeyearpar{hsu2022u} & Noisy & Noisy & - & - & 4.1 & - & - & - & - & - & - & - & - & - & - & - & - & - & - & - & 1.2 \\
\midrule
\multirow{4}{*}{AV-HuBERT~\citeyearpar{shi2022robust}} & \multirow{2}{*}{Clean} & Clean & 72.6 & 30.9 & 9.8 & 2.9 & 2.1 & 23.7 & 93.4 & 71.6 & 22.1 & 6.1 & 2.7 & 39.2 & 24.1 & 10.9 & 3.6 & 2.4 & 1.9 & 8.6 & 1.42 \\
& & Noisy & 30.0 & 15.2 & 5.9 & 2.7 & 1.9 & 11.1 & 15.9 & 7.5 & 3.9 & 2.4 & 1.9 & 6.3 & 12.1 & 5.9 & 3.1 & 2.2 & 1.8 & 5.0 & 1.40 \\
\cline{2-22}
& \multirow{2}{*}{Noisy} & Clean & 39.4 & 14.5 & 5.2 & 2.7 & 2.0 & 12.8 & 18.8 & 5.1 & 3.1 & 2.3 & 1.9 & 6.2 & 11.4 & 5.0 & 2.8 & 2.2 & 1.8 & 4.6 & 1.54 \\
& & Noisy & 28.4 & 13.4 & 5.0 & 2.6 & 1.9 & 10.3 & 11.4 & 4.6 & 2.9 & 2.2 & 1.8 & 4.6 & 9.7 & 4.7 & 2.5 & 1.9 & 1.8 & 4.1 & 1.40 \\
\midrule
\multirow{4}{*}{UniVPM (ours)} & \multirow{2}{*}{Clean} & Clean & \cellhl37.5 & \cellhl17.1 & \cellhl6.9 & \cellhl2.6 & \cellhl1.9 & \cellhl13.2 & \cellhl20.4 & \cellhl9.6 & \cellhl4.9 & \cellhl3.6 & \cellhl2.3 & \cellhl8.2 & \cellhl14.2 & \cellhl6.8 & \cellhl3.1 & \cellhl2.1 & \cellhl1.8 & \cellhl5.6 & 1.31 \\
& & Noisy & 28.1 & 13.8 & 5.1 & 2.2 & 1.7 & 10.2 & 14.5 & 6.7 & 3.3 & 2.1 & 1.7 & 5.7 & 10.7 & 5.2 & 2.7 & 1.9 & 1.6 & 4.4 & 1.22 \\
\cline{2-22}
& \multirow{2}{*}{Noisy} & Clean & 32.6 & 12.6 & 4.4 & 2.3 & 1.7 & 10.7 & 17.0 & 4.4 & 2.7 & 2.1 & 1.6 & 5.6 & 10.1 & 4.3 & 2.4 & 1.9 & 1.6 & 4.1 & 1.25 \\
& & Noisy & \textbf{26.8} & \textbf{12.1} & \textbf{4.0} & \textbf{2.1} & \textbf{1.6} & \textbf{9.3} & \textbf{10.4} & \textbf{4.1} & \textbf{2.5} & \textbf{2.0} & \textbf{1.6} & \textbf{4.1} & \textbf{8.7} & \textbf{4.1} & \textbf{2.1} & \textbf{1.7} & \textbf{1.5} & \textbf{3.6} & \textbf{1.18} \\
\bottomrule
\end{tabular}}
\vspace{-0.15cm}
\caption{WER (\%) of proposed UniVPM and prior works on LRS3 benchmark.
``PT Type'' / ``FT Type" denote pre-training / finetuning data type.
``SNR'' is signal-to-noise ratio.
All the noisy data contains MUSAN~\citeyearpar{snyder2015musan} noise.}
\label{table1}
\end{table*}

\begin{table*}[t]
\centering
\resizebox{1.0\textwidth}{!}{
\begin{tabular}{ccc|cccccc|cccccc|cccccc|c}
\toprule
\multirow{2}{*}{Method} & PT & FT & \multicolumn{6}{c|}{Babble, SNR (dB) =} & \multicolumn{6}{c|}{Speech, SNR (dB) =} & \multicolumn{6}{c|}{Music + Natural, SNR (dB) =} & Clean \\ 
& Type & Type & -10 & -5 & 0 & 5 & 10 & avg & -10 & -5 & 0 & 5 & 10 & avg & -10 & -5 & 0 & 5 & 10 & avg & $\infty$  \\ 
\midrule
TM-seq2seq~\citeyearpar{afouras2018deep} & - & Noisy & - & - & - & - & - & - & - & - & - & - & - & - & - & - & - & - & - & - & 8.5 \\
Hyb-RNN~\citeyearpar{petridis2018audio} & - & Noisy & - & - & - & - & - & - & - & - & - & - & - & - & - & - & - & - & - & - & 7.0 \\
LF-MMI TDNN~\citeyearpar{yu2020audio} & - & Clean & - & - & - & - & - & - & - & - & - & - & - & - & - & - & - & - & - & - & 5.9 \\
Hyb-AVSR~\citeyearpar{ma2021end} & - & Noisy & - & - & - & - & - & - & - & - & - & - & - & - & - & - & - & - & - & - & 3.7 \\
MoCo+w2v2~\citeyearpar{pan2022leveraging} & - & Noisy & - & - & - & - & - & - & - & - & - & - & - & - & - & - & - & - & - & - & 2.6 \\
\midrule
\multirow{4}{*}{AV-HuBERT~\citeyearpar{shi2022robust}} & \multirow{2}{*}{Clean} & Clean & 65.2 & 33.6 & 10.9 & 5.6 & 3.8 & 23.8 & 88.2 & 57.8 & 20.6 & 7.5 & 4.0 & 35.6 & 27.3 & 13.3 & 6.7 & 4.0 & 3.4 & 10.9 & 2.57 \\
& & Noisy & 33.2 & 16.3 & 7.6 & 4.6 & 3.7 & 13.1 & 14.9 & 9.5 & 6.2 & 4.5 & 3.8 & 7.8 & 13.9 & 9.0 & 4.9 & 3.9 & 3.2 & 7.0 & 2.38 \\
\cline{2-22}
& \multirow{2}{*}{Noisy} & Clean & 36.9 & 18.6 & 8.1 & 4.8 & 3.5 & 14.4 & 24.6 & 9.7 & 4.8 & 3.6 & 3.4 & 9.2 & 15.2 & 8.4 & 5.1 & 3.8 & 3.1 & 7.1 & 2.44 \\
& & Noisy & 32.7 & 14.9 & 6.4 & 4.5 & 3.4 & 12.4 & 9.0 & 5.9 & 3.9 & 3.5 & 3.0 & 5.1 & 12.5 & 6.0 & 4.4 & 3.5 & 3.0 & 5.9 & 2.33 \\
\midrule
\multirow{4}{*}{UniVPM (ours)} & \multirow{2}{*}{Clean} & Clean & \cellhl38.3 & \cellhl19.0 & \cellhl9.2 & \cellhl5.0 & \cellhl3.5 & \cellhl15.0 & \cellhl21.1 & \cellhl12.2 & \cellhl7.8 & \cellhl5.4 & \cellhl3.9 & \cellhl10.1 & \cellhl16.3 & \cellhl10.4 & \cellhl5.6 & \cellhl3.6 & \cellhl3.2 & \cellhl7.8 & 2.30 \\
& & Noisy & 30.4 & 14.4 & 6.6 & \textbf{4.1} & 3.4 & 11.8 & 12.4 & 8.3 & 5.5 & 4.2 & 3.6 & 6.8 & 12.4 & 7.9 & 4.3 & 3.4 & 3.0 & 6.2 & 2.17 \\
\cline{2-22}
& \multirow{2}{*}{Noisy} & Clean & 33.7 & 16.2 & 6.7 & 4.2 & \textbf{3.2} & 12.8 & 19.8 & 7.6 & 4.0 & 3.2 & 3.1 & 7.5 & 13.4 & 7.3 & 4.5 & 3.4 & 2.9 & 6.3 & 2.24 \\
& & Noisy & \textbf{30.1} & \textbf{13.7} & \textbf{5.7} & \textbf{4.1} & \textbf{3.2} & \textbf{11.4} & \textbf{7.5} & \textbf{5.1} & \textbf{3.4} & \textbf{3.1} & \textbf{2.8} & \textbf{4.4} & \textbf{10.9} & \textbf{5.0} & \textbf{3.8} & \textbf{3.1} & \textbf{2.8} & \textbf{5.1} & \textbf{2.16} \\
\bottomrule
\end{tabular}}
\vspace{-0.15cm}
\caption{WER (\%) of proposed UniVPM and prior works on LRS2 benchmark.}
\label{table2}
\vspace{-0.3cm}
\end{table*}

\section{Experiments}
\label{sec:exp}

\subsection{Experimental Setup}
\label{ssec:setup}
\label{sssec:datasets}
\noindent\textbf{Datasets.} Our experiments are conducted on two large-scale public datasets, LRS3~\citep{afouras2018lrs3} and LRS2~\citep{chung2017lip}.
LRS3 dataset collects 433 hours of transcribed English videos from TED \& TEDx talks. 
LRS2 contains 224 hours of video speech from BBC programs.

\label{sssec:config_baselines}
\noindent\textbf{Configurations and Baselines.} 
The proposed UniVPM is implemented based on AV-HuBERT with similar configurations, which are detailed in \S\ref{assec:model_config}.
We also select some mainstream AVSR approaches as baselines for comparison, \textit{e.g.}, u-HuBERT~\citep{hsu2022u}, and details are presented in \S\ref{assec:baselines}.



\begin{table*}[t]
\centering
\resizebox{1.0\textwidth}{!}{
\begin{tabular}{ccc|ccccc|ccccc|ccccc|ccccc}
\toprule
\multirow{2}{*}{Method} & PT & FT & \multicolumn{5}{c|}{Meeting, SNR (dB) =} & \multicolumn{5}{c|}{Cafe, SNR (dB) =} & \multicolumn{5}{c|}{Resto, SNR (dB) =} & \multicolumn{5}{c}{Station, SNR (dB) =} \\ 
& Type & Type & -10 & -5 & 0 & 5 & avg & -10 & -5 & 0 & 5 & avg & -10 & -5 & 0 & 5 & avg & -10 & -5 & 0 & 5 & avg  \\
\midrule
\multicolumn{23}{c}{\cellcolor[HTML]{EBEBEB} \emph{Finetuned on DEMAND Noise}}  \vspace{0.05cm} \\
\multirow{4}{*}{AV-HuBERT~\citeyearpar{shi2022robust}} & \multirow{2}{*}{Clean} & Clean & 33.2 & 11.7 & 4.3 & 3.1 & 13.1 & 26.0 & 8.5 & 2.9 & 2.0 & 9.9 & 63.5 & 30.4 & 11.0 & 3.9 & 27.2 & 20.1 & 7.0 & 4.7 & 2.5 & 8.6 \\
& & Noisy & 10.6 & 5.2 & 2.9 & 2.5 & 5.3 & 10.1 & 4.3 & 2.3 & 1.8 & 4.6 & 27.8 & 14.4 & 4.9 & 2.6 & 12.4 & 7.6 & 4.5 & 2.9 & 2.0 & 4.3 \\
\cline{2-23}
& \multirow{2}{*}{Noisy} & Clean & 17.7 & 7.1 & 4.0 & 2.9 & 7.9 & 16.0 & 5.8 & 2.7 & 1.9 & 6.6 & 49.5 & 19.5 & 6.2 & 3.1 & 19.6 & 11.8 & 5.9 & 3.7 & 2.2 & 5.9 \\
& & Noisy & 10.2 & 4.8 & 2.7 & 2.4 & 5.0 & 9.4 & 4.0 & 2.2 & 1.8 & 4.4 & 23.5 & 13.2 & 4.4 & \textbf{2.4} & 10.9 & 7.2 & \textbf{4.3} & 2.9 & 1.8 & 4.1 \\
\midrule
\multicolumn{23}{c}{\cellcolor[HTML]{EBEBEB} \emph{Finetuned on MUSAN Noise}}  \vspace{0.05cm} \\
\multirow{4}{*}{AV-HuBERT~\citeyearpar{shi2022robust}} & \multirow{2}{*}{Clean} & Clean & 33.2 & 11.7 & 4.3 & 3.1 & 13.1 & 26.0 & 8.5 & 2.9 & 2.0 & 9.9 & 63.5 & 30.4 & 11.0 & 3.9 & 27.2 & 20.1 & 7.0 & 4.7 & 2.5 & 8.6 \\
& & Noisy & 13.9 & 6.3 & 3.3 & 2.8 & 6.6 & 13.6 & 5.1 & 2.6 & 1.9 & 5.8 & 36.1 & 17.5 & 5.7 & 2.9 & 15.6 & 9.9 & 5.3 & 3.5 & 2.1 & 5.2 \\
\cline{2-23}
& \multirow{2}{*}{Noisy} & Clean & 17.7 & 7.1 & 4.0 & 2.9 & 7.9 & 16.0 & 5.8 & 2.7 & 1.9 & 6.6 & 49.5 & 19.5 & 6.2 & 3.1 & 19.6 & 11.8 & 5.9 & 3.7 & 2.2 & 5.9 \\
& & Noisy & 13.2 & 5.5 & 3.2 & 2.7 & 6.2 & 12.4 & 4.8 & 2.3 & 1.8 & 5.3 & 33.7 & 16.1 & 5.1 & 2.6 & 14.4 & 9.8 & 5.1 & 3.5 & 1.9 & 5.1 \\
\midrule
\multirow{4}{*}{UniVPM (ours)} & \multirow{2}{*}{Clean} & Clean & \cellhl12.8 & \cellhl5.3 & \cellhl3.1 & \cellhl2.7 & \cellhl6.0 & \cellhl12.1 & \cellhl4.9 & \cellhl2.3 & \cellhl1.7 & \cellhl5.3 & \cellhl32.8 & \cellhl15.8 & \cellhl5.0 & \cellhl2.8 & \cellhl14.1 & \cellhl9.5 & \cellhl5.0 & \cellhl3.6 & \cellhl2.1 & \cellhl5.1 \\
& & Noisy & 10.0 & 4.7 & 2.7 & 2.4 & 5.0 & 9.6 & 4.0 & 2.2 & \textbf{1.6} & 4.4 & 24.9 & 13.3 & 4.7 & 2.6 & 11.4 & 7.0 & \textbf{4.3} & 2.9 & 1.8 & 4.0 \\
\cline{2-23}
& \multirow{2}{*}{Noisy} & Clean & 11.9 & 5.1 & 3.0 & 2.6 & 5.7 & 10.8 & 4.6 & 2.2 & 1.7 & 4.8 & 27.4 & 14.8 & 4.9 & 2.6 & 12.4 & 8.3 & 4.7 & 3.2 & 1.8 & 4.5 \\
& & Noisy & \textbf{9.7} & \textbf{4.6} & \textbf{2.6} & \textbf{2.3} & \textbf{4.8} & \textbf{9.0} & \textbf{3.8} & \textbf{2.1} & \textbf{1.6} & \textbf{4.1} & \textbf{22.6} & \textbf{12.9} & \textbf{4.3} & \textbf{2.4} & \textbf{10.6} & \textbf{6.9} & \textbf{4.3} & \textbf{2.8} & \textbf{1.7} & \textbf{3.9} \\
\bottomrule
\end{tabular}}
\vspace{-0.2cm}
\caption{WER (\%) on unseen testing noises with LRS3 benchmark.
Testing noises ``\texttt{Meeting}'', ``\texttt{Cafe}'', ``\texttt{Resto}'' and ``\texttt{Station}'' are from DEMAND dataset~\citeyearpar{thiemann2013diverse}.
Pre-training noise are from MUSAN dataset.}
\label{table3}
\vspace{-0.2cm}
\end{table*}

\subsection{Main Results}
\label{ssec:main_results}

\noindent\textbf{Audio-Visual Speech Recognition.}
Table~\ref{table1} compares the AVSR performance of our UniVPM with prior methods on LRS3 benchmark.
Under clean training data, the proposed UniVPM (in purple shades) significantly outperforms AV-HuBERT baseline, and it achieves comparable performance to the AV-HuBERT trained on noisy data, where the restored clean audio plays the key role and implements our original motivation of unsupervised noise adaptation.
Based on that, available noisy training data further improves the robustness\footnote{Noisy training pipeline of UniVPM is shown in Fig.~\ref{fig9}.}, where our best results achieve new state-of-the-art in various noisy as well as clean conditions.
Furthermore, we can also observe similar results on LRS2 dataset as shown in Table~\ref{table2}.

\label{sssec:unseen_noise}
Table~\ref{table3} further compares the performance of UniVPM with AV-HuBERT on unseen testing noises, which are sampled from DEMAND~\citep{thiemann2013diverse} dataset.
First, when AV-HuBERT is finetuned and tested both on DEMAND noise, good WER performance can be achieved.
However, if it is finetuned on MUSAN noise and then tested on unseen DEMAND noise, the performance would degrade a lot.
In comparison, our UniVPM finetuned on clean data (purple shades) achieves significant improvement and surpasses the AV-HuBERT finetuned on MUSAN noise, which further verifies the strong generality of our model.
Furthermore, when finetuned on MUSAN noise, our UniVPM even outperforms the AV-HuBERT finetuned on in-domain DEMAND noise, which highlights the superiority of our approach on unseen test noises.

\begin{table}[t]
\centering
\resizebox{1.0\columnwidth}{!}{
\begin{tabular}{cccc|c}
\toprule
\multirow{2}{*}{Method} & Finetune & Unlabeled & Labeled & \multirow{2}{*}{WER (\%)} \\ 
& Mode & Data (hrs) & Data (hrs) & \\ 
\midrule
TM-seq2seq~\citeyearpar{afouras2018deep} & AV & - & 1,519 & 58.9 \\
EG-seq2seq~\citeyearpar{xu2020discriminative} & AV & - & 590 & 57.8 \\
Hyb-AVSR~\citeyearpar{ma2021end} & AV & - & 590 & 43.3 \\
RNN-T~\citeyearpar{makino2019recurrent} & AV & - & 31,000 & 33.6 \\
Distill-PT~\citeyearpar{ma2022visual} & V & 1,021 & 438 & 31.5 \\
u-HuBERT~\citeyearpar{hsu2022u} & AV & 2,211 & 433 & 27.2 \\
\midrule
\multirow{2}{*}{AV-HuBERT~\citeyearpar{shi2022learning}} & AV & 1,759 & 433 & 34.7 \\
& V & 1,759 & 433 & 28.6 \\
\quad + Self-Training & V & 1,759 & 433 & 26.9 \\
\midrule
UniVPM (ours) & AV & 1,759 & 433 & \textbf{26.7} \\
\bottomrule
\end{tabular}}
\vspace{-0.1cm}
\caption{WER (\%) results of visual speech recognition (VSR) on LRS3 benchmark. ``Finetune Mode'' denotes the input modality during finetuning stage.}
\label{table4}
\vspace{-0.5cm}
\end{table}

\vspace{0.05cm}
\noindent\textbf{Visual Speech Recognition.}
To further verify the effectiveness of UniVPM, we evaluate its VSR performance by discarding the input audio modality during inference, as shown in Table~\ref{table4}.
In this case, with restored clean audio from lip movements, the proposed UniVPM significantly outperforms AV-HuBERT baseline (34.7\%$\rightarrow$26.7\%).
Although the visual-only training and self-training strategies improve AV-HuBERT's results, our UniVPM still defines new state-of-the-art on LRS3 benchmark.

\subsection{Ablation Study}
\label{ssec:ablation}
Table~\ref{table5} presents the ablation study of components in UniVPM.
The four parts of ablation are independent, \textit{i.e.}, each study is conducted where other three components are kept same as full UniVPM.

\begin{table}[t]
\centering
\resizebox{1.0\columnwidth}{!}{
\begin{tabular}{c|cccc|c|c}
\toprule
Method & B & S & M & N & Clean & VSR \\
\midrule
AV-HuBERT~\citeyearpar{shi2022robust} & 23.7 & 39.2 & 10.7 & 6.4 & 1.42 & 34.7 \\
\midrule
\multicolumn{7}{c}{\cellcolor[HTML]{EBEBEB} \emph{Effectiveness of Online Balanced Clustering}}  \vspace{0.05cm}   \\
Memory Network~\citeyearpar{kim2022distinguishing} & 20.6 & 29.5 & 9.2 & 6.1 & 1.39 & 32.0 \\
Online Clustering & 19.3 & 22.9 & 8.7 & 5.9 & 1.37 & 31.2 \\
Online Balanced Clustering & \textbf{13.2} & \textbf{8.2} & \textbf{6.1} & \textbf{5.1} & \textbf{1.31} & \textbf{26.7} \\
\midrule
\multicolumn{7}{c}{\cellcolor[HTML]{EBEBEB} \emph{Effectiveness of AMIE}}  \vspace{0.05cm}   \\
None & 22.3 & 35.4 & 10.4 & 6.0 & 1.39 & 31.8 \\
Contrastive Learning & 21.5 & 29.2 & 9.7 & 5.8 & 1.37 & 30.1 \\
MINE~\citeyearpar{belghazi2018mutual} & 18.6 & 20.1 & 8.3 & 5.5 & 1.34 & 28.8 \\
AMIE w/o Adv. Learning & 17.0 & 17.9 & 7.7 & 5.4 & 1.33 & 28.2 \\
AMIE & \textbf{13.2} & \textbf{8.2} & \textbf{6.1} & \textbf{5.1} & \textbf{1.31} & \textbf{26.7} \\
\midrule
\multicolumn{7}{c}{\cellcolor[HTML]{EBEBEB} \emph{Analysis of Adversarial Learning}}  \vspace{0.05cm} \\
$\mathcal{L}_{G\hspace{-0.02cm}A\hspace{-0.02cm}N}$ w/ $I(s_v, s_a)$ & 15.4 & 14.6 & 7.2 & 5.3 & 1.32 & 27.8 \\
$\mathcal{L}_{G\hspace{-0.02cm}A\hspace{-0.02cm}N}$ w/ $I(f_v, \hat{f}_a)$ & 17.7 & 22.0 & 8.8 & 5.6 & 1.36 & 29.2 \\
$\mathcal{L}_{G\hspace{-0.02cm}A\hspace{-0.02cm}N}$ w/ $I(s_v, s_a) + I(f_v, \hat{f}_a)$ & \textbf{13.2} & \textbf{8.2} & \textbf{6.1} & \textbf{5.1} & \textbf{1.31} & \textbf{26.7} \\
\midrule
\multicolumn{7}{c}{\cellcolor[HTML]{EBEBEB} \emph{Analysis of Regularization}}  \vspace{0.05cm}   \\
None & 17.2 & 20.4 & 8.0 & 5.7 & 1.36 & 30.3 \\
UniVPM w/ $\mathcal{L}_{rec}$ & 14.3 & 11.5 & 6.5 & 5.3 & 1.33 & 27.4 \\
UniVPM w/ $\mathcal{L}_{var}$ & 15.6 & 14.6 & 7.2 & 5.4 & 1.33 & 28.5 \\
UniVPM w/ $\mathcal{L}_{rec} + \mathcal{L}_{var}$ & \textbf{13.2} & \textbf{8.2} & \textbf{6.1} & \textbf{5.1} & \textbf{1.31} & \textbf{26.7} \\
\bottomrule
\end{tabular}}
\vspace{-0.1cm}
\caption{Ablation study.
`B', `S', `M', `N' denote average-SNR results on four MUSAN noises in Table~\ref{table1}.
``Adv.'' denotes ``Adversarial''.
The four ablations are all based on full UniVPM and independent with each other.}
\label{table5}
\vspace{-0.4cm}
\end{table}

\begin{figure*}[t]
\centering
    \includegraphics[width=0.81\textwidth]{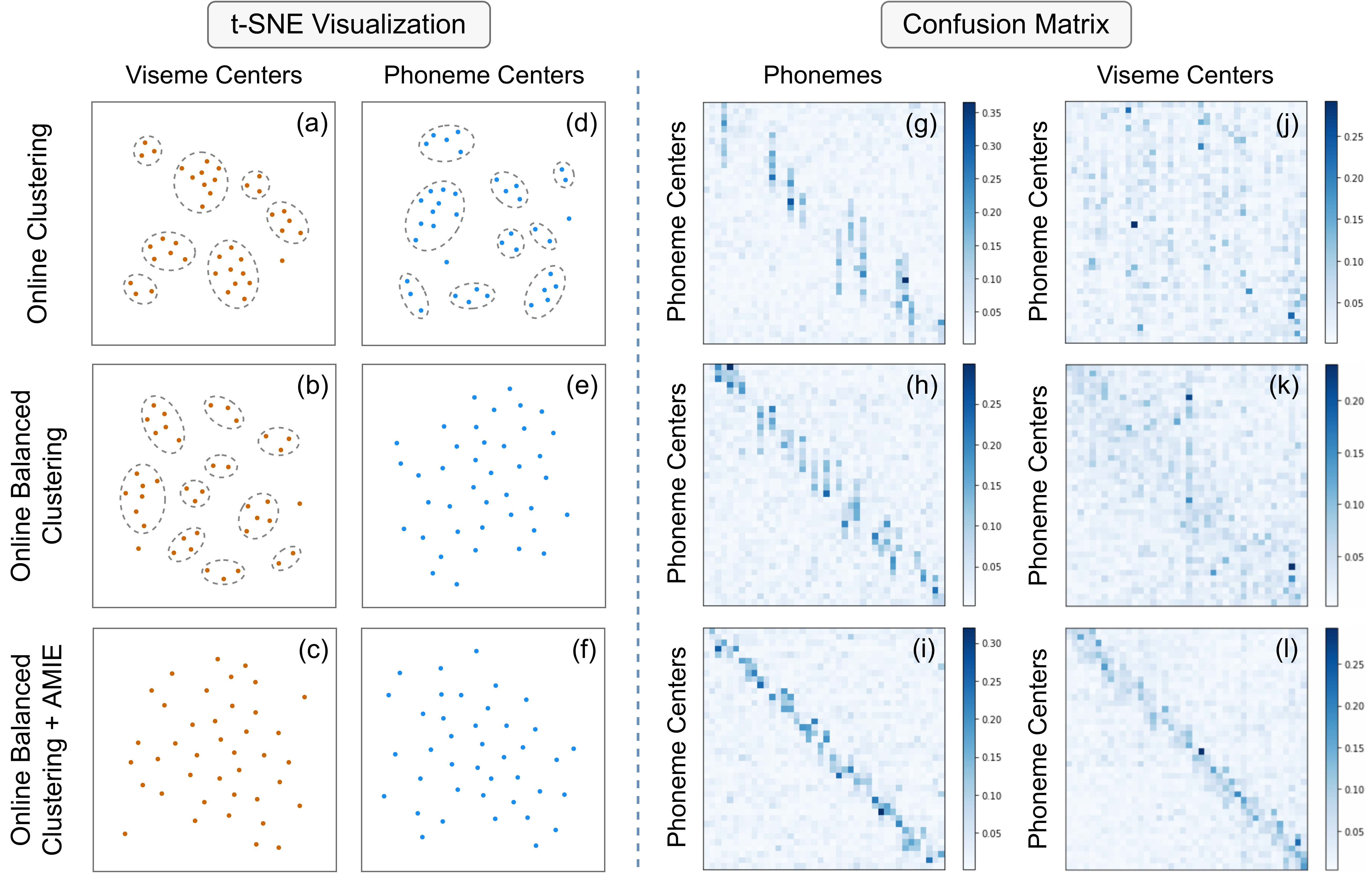}
    \vspace{-0.2cm}
    \caption{Left panel: t-SNE visualization of clustered viseme and phoneme centers (ellipses highlight the undesirably gathered centers). 
    Right panel: confusion matrix of phoneme matching and viseme-phoneme mapping.
    In (g)-(i), the vertical axis indicates phoneme center IDs and the horizontal axis indicates real phonemes predicted by pre-trained model~\citep{phy22phoneme}, while in (j)-(l) the horizontal axis indicates viseme center IDs.}\label{fig5}
    \vspace{-0.45cm}
\end{figure*}

\vspace{0.05cm}
\noindent\textbf{Effect of Online Balanced Clustering.}
In UniVPM, our online clustering baseline outperforms the memory network with learnable embeddings, indicating the superiority of clustering technique in representing visemes and phonemes.
Based on that, our proposed online balanced clustering achieves significant improvement by modeling all the visemes and phonemes equally without over-fitting, which is further shown in Fig.~\ref{fig5}.

\vspace{0.05cm}
\noindent\textbf{Effect of AMIE.}
As presented in Table~\ref{table5}, AMIE plays the key role in the promising performance of UniVPM by constructing strong viseme-phoneme mapping.
As a comparison, the contrastive learning baseline only provides limited improvement, and MINE performs better by maximizing the estimated MI between visemes and phonemes.
Based on that, our proposed AMIE introduces JS representation to stabilize system optimization, which improves performance but still suffers from the ambiguity of homophenes.
To this end, our adversarial learning approach achieves further improvement by constructing strict viseme-phoneme mapping without ambiguity, as shown in Fig.~\ref{fig8}.

\vspace{0.05cm}
\noindent\textbf{Analysis of Adversarial Learning.}
As illustrated in Eq.~\ref{eq11}, there are two key components in adversarial learning, \textit{i.e.}, $I(s_v, s_a)$ that constructs viseme-phoneme mapping and $I(f_v, \hat{f}_a)$ that supervises the quality of restored clean audio.
Results in Table~\ref{table5} indicate that viseme-phoneme mapping is the most important, and the supervision on restored clean audio also improves the AVSR performance.

\vspace{0.02cm}
\noindent\textbf{Analysis of Regularization.}
According to Eq.~\ref{eq12}, $\mathcal{L}_{rec}$ and $\mathcal{L}_{var}$ are two auxiliary terms for regularization, where the former supervises the quality of restored audio, and the latter disperses clustered viseme and phoneme centers to ease their mapping construction.
Both of them are proved with positive contributions to the gains of performance.

\vspace{0.02cm}
\noindent\textbf{Visualizations.}
Fig.~\ref{fig5} presents t-SNE visualization and confusion matrixes to further verify the effectiveness of UniVPM.
First, the online clustering baseline generates gathered viseme and phoneme centers due to over-fitting, where only several majority phonemes are modeled as shown in (g).
Our proposed online balanced clustering alleviates such over-fitting issue and generates separated phoneme centers, which can cover most of the real phonemes as illustrated in (h).
However, we can still observe gathered viseme centers due to homophenes, and the ambiguity of viseme-phoneme mapping is also shown in (k).
To this end, our proposed AMIE effectively alleviates the ambiguity of homophenes thanks to the strong distinguishing ability of adversarial learning, which constructs strict viseme-phoneme mapping in (l).
Meanwhile, we also observe dispersed viseme centers in (c), which can distinguish the same visemes that correspond to different phonemes.
In addition, real phonemes are also better modeled by clustering as shown in (i).

\begin{table}[t]
\centering
\vspace{0.1cm}
\resizebox{1.0\columnwidth}{!}{
\begin{tabular}{l|c|c}
\toprule
\multicolumn{1}{c|}{\multirow{2}{*}{Method}} & VSR & Phoneme \\
& WER (\%) & Match Acc. (\%) \\
\midrule
AV-HuBERT~\citeyearpar{shi2022learning} & 34.7 & - \\
\quad + Online Clustering & 33.5 & 14.2 \\
\quad + Online Balanced Clustering & 31.8 & 31.0 \\
\quad\quad + AMIE (UniVPM) & \textbf{26.7} & \textbf{67.5} \\
\bottomrule
\end{tabular}}
\vspace{-0.15cm}
\caption{Evaluation of restored clean audio in terms of phoneme match accuracy on LRS3 test set.
It is calculated with predicted phonemes for restored audio and real clean audio by pre-trained model~\citep{phy22phoneme}.}
\label{table6}
\vspace{-0.5cm}
\end{table}

\vspace{0.02cm}
\noindent\textbf{Evaluation of Modality Transfer.}
Table~\ref{table6} further reports phoneme match accuracy to evaluate the quality of restored clean audio.
We observe that online clustering baseline can hardly restore correct phonemes, and the proposed online balanced clustering improves the accuracy but still limited by the ambiguity of homophenes.
Furthermore, our proposed AMIE significantly improves the quality of restored clean audio with strict viseme-phoneme mapping, which also yields better VSR result.

\section{Conclusion}
\label{sec:conclusion}
In this paper, we propose UniVPM, a general robust AVSR approach motivated from visual modality via unsupervised noise adaptation.
UniVPM constructs universal viseme-phoneme mapping to implement modality transfer, which can restore clean audio from visual signals to enable speech recognition under any noises.
Experiments on public benchmarks show that UniVPM achieves state-of-the-art under various noisy as well as clean conditions.
Further analysis also verifies its effectiveness on VSR task.

\section*{Limitations}
We state two points of limitations and future work in this section.
First, the UniVPM combines both restored clean audio and original input audio for downstream speech recognition, while without any trade-off to weight them.
For example, under extremely noisy conditions the restored clean audio plays a more important role, while in less noisy scenarios the original audio may provide more valid information.
Some weighting strategies to select the most effective audio information could benefit the downstream speech recognition.
Second, the proposed clustering and viseme-phoneme mapping are actually unsupervised schemes, so that it could be promising to extend our UniVPM to the popular self-supervised learning framework, in order to make full use of the abundant unlabeled data.

\section*{Ethics Statement}
All the data used in this paper are publicly available and are used under the following licenses: the Creative Commons BY-NC-ND 4.0 License and Creative Commons Attribution 4.0 International License, the TED Terms of Use, the YouTube's Terms of Service, and the BBC’s Terms of Use.
The data is collected from TED and BBC and contain thousands of speakers from a wide range of races.
To protect the anonymity, only mouth area of speaker is visualized wherever used in the paper.

\section*{Acknowledgements}
This research is supported by KLASS Engineering \& Solutions Pte Ltd and the National Research Foundation, Singapore under its AI Singapore Programme (AISG Award No.: AISG2-100E-2023-103).
The computational work for this article was partially performed on resources of the National Supercomputing Centre, Singapore (\url{https://www.nscc.sg}).


\bibliographystyle{acl_natbib}

\appendix

\section{Supplementary Experimental Analysis}
\label{asec:supple_exp_analysis}

\begin{table*}[t]
\centering
\resizebox{1.0\textwidth}{!}{
\begin{tabular}{ccc|cccccc|cccccc|cccccc|c}
\toprule
\multirow{2}{*}{Mode} & PT & FT & \multicolumn{6}{c|}{Babble, SNR (dB) =} & \multicolumn{6}{c|}{Speech, SNR (dB) =} & \multicolumn{6}{c|}{Music + Natural, SNR (dB) =} & Clean \\ 
& Type & Type & -10 & -5 & 0 & 5 & 10 & avg & -10 & -5 & 0 & 5 & 10 & avg & -10 & -5 & 0 & 5 & 10 & avg & $\infty$  \\ 
\midrule
\multirow{4}{*}{A} & \multirow{2}{*}{Clean} & Clean & 99.3 & 89.6 & 43.9 & 11.0 & 3.7 & 49.5 & 102.5 & 93.8 & 63.5 & 24.1 & 10.7 & 58.9 & 58.6 & 35.9 & 13.9 & 5.4 & 2.6 & 23.3 & 1.55 \\
& & Noisy & 98.2 & 65.6 & 17.0 & 5.3 & 2.7 & 37.8 & 94.3 & 73.8 & 46.3 & 22.9 & 9.7 & 49.4 & 43.4 & 18.0 & 6.5 & 3.2 & 2.1 & 14.6 & 1.50 \\
\cline{2-22}
& \multirow{2}{*}{Noisy} & Clean & 98.3 & 77.6 & 23.0 & 7.3 & 2.9 & 41.8 & 87.3 & 62.9 & 41.0 & 22.2 & 8.9 & 44.5 & 43.4 & 19.3 & 7.1 & 3.4 & 2.5 & 15.1 & 1.62 \\
& & Noisy & 97.5 & 62.3 & 15.7 & 5.1 & 2.6 & 36.6 & 81.7 & 56.2 & 37.3 & 19.0 & 8.3 & 40.5 & 38.7 & 15.1 & 5.7 & 3.1 & 2.3 & 13.0 & 1.60 \\
\midrule
\multirow{4}{*}{AV} & \multirow{2}{*}{Clean} & Clean & \cellhl72.6 & \cellhl30.9 & \cellhl9.8 & \cellhl2.9 & \cellhl2.1 & \cellhl23.7 & \cellhl93.4 & \cellhl71.6 & \cellhl22.1 & \cellhl6.1 & \cellhl2.7 & \cellhl39.2 & \cellhl24.1 & \cellhl10.9 & \cellhl3.6 & \cellhl2.4 & \cellhl1.9 & \cellhl8.6 & 1.42 \\
& & Noisy & 30.0 & 15.2 & 5.9 & 2.7 & 1.9 & 11.1 & 15.9 & 7.5 & 3.9 & 2.4 & 1.9 & 6.3 & 12.1 & 5.9 & 3.1 & 2.2 & 1.8 & 5.0 & 1.40 \\
\cline{2-22}
& \multirow{2}{*}{Noisy} & Clean & 39.4 & 14.5 & 5.2 & 2.7 & 2.0 & 12.8 & 18.8 & 5.1 & 3.1 & 2.3 & 1.9 & 6.2 & 11.4 & 5.0 & 2.8 & 2.2 & 1.8 & 4.6 & 1.54 \\
& & Noisy & 28.4 & 13.4 & 5.0 & 2.6 & 1.9 & 10.3 & 11.4 & 4.6 & 2.9 & 2.2 & 1.8 & 4.6 & 9.7 & 4.7 & 2.5 & 1.9 & 1.8 & 4.1 & 1.40 \\
\bottomrule
\end{tabular}}
\vspace{-0.1cm}
\caption{WER (\%) of AV-HuBERT on LRS3 benchmark.
``Mode'' denotes the input modality during both finetuning and inference stages, ``PT Type" denotes the pre-training data type, ``FT Type" denotes the finetuning data type, and ``avg'' denotes the average performance on all SNRs.}
\label{table7}
\vspace{-0.4cm}
\end{table*}

\subsection{Analysis of the Noise-Robustness of AVSR}
\label{assec:analysis_robustness_avsr}
Table~\ref{table7} presents the performance of AV-HuBERT to analyze the noise-robustness of AVSR system.
First, as the original motivation of AVSR, the visual modality significantly improves the audio-only speech recognition performance under various noisy as well as clean testing conditions, especially the low-SNR environments.
However, most existing efforts still focus on audio modality to improve robustness considering its dominance in AVSR task. The reason is the inherent information insufficiency of visual modality to represent speech content.
Mainstream approaches introduce noise adaptation techniques~\cite{hu2022interactive,hu2022dual,chen2022self,chen2023metric,chen2023unsupervised,hu2023gradient,hu2023unifying,hu2023wav2code,zhu2023joint,zhu2023robust} to strengthen robustness, where most of them leverage noise-corrupted data to improve network training~\citep{afouras2018deep,ma2021end,pan2022leveraging,shi2022robust,hsu2022u,chen2022leveraging,hu2023cross,zhu2023vatlm}.
As shown in Table~\ref{table7}, available noisy training data significantly improves the AVSR performance in different testing conditions.
However, this strategy is usually faced with two practical challenges.
First, it requires abundant labeled noisy audio-visual training data, which is not always available in some real-world scenarios~\citep{meng2017unsupervised,long2017domain,lin2021unsupervised,chen2022noise}.
For instance, in scenarios like theatre, it is valuable to develop a AVSR system but costly to obtain sufficient training data.
Second, as it is impossible to cover all the real-world noises in training data, when some unseen noises appear in practical testing scenarios, the well-trained model may not perform well as shown in Table~\ref{table3}, resulting in less optimal model generality~\citep{meng2017unsupervised}.
Above two challenges motivate this work.
With unsupervised noise adaptation investigated on visual modality, our proposed UniVPM improves the AVSR performance under clean training data to a comparable level to the state-of-the-art AV-HuBERT trained on noisy data in various noisy as well as clean testing conditions, as shown in Table~\ref{table1},~\ref{table2}, and~\ref{table3}.
Moreover, available noisy training data can further improve the robustness of UniVPM and yield new state-of-the-arts on both LRS3 and LRS2 benchmarks.

\subsection{Analysis of Limited In-domain Noisy Audio-Visual Data}
\label{assec:analysis_limited_data}
According to \S\ref{sec:intro} and \S\ref{assec:analysis_robustness_avsr}, the first challenge of audio modality-based robust AVSR is the limited in-domain noisy audio-visual data, which leads to domain mismatch between training and testing~\citep{meng2017unsupervised,long2017domain,lin2021unsupervised,chen2020data,chen2022noise}.
Actually there are two methods of obtaining such data, \textit{i.e.}, collection and simulation.
First, we can collect and transcribe amounts of noisy audio-visual data under real-world scenarios, but that is extremely time-consuming and laborious, and to our best knowledge there is currently no such public dataset.
Second, as there is sufficient clean transcribed audio-visual data~\citep{afouras2018lrs3,chung2017lip}, we can collect in-domain noise to simulate noisy audio-visual data.
However, this data augmentation method can only partially alleviate but not resolve the domain mismatch problem~\citep{zhang2022monoaural}.
What is worse, the in-domain noise data is also not always available in all the real-world scenarios~\citep{meng2017unsupervised,long2017domain,chen2020data,chen2022noise}.

As presented in Table~\ref{table1}, in case of no available in-domain noise, our UniVPM achieves comparable performance to previous state-of-the-art trained on in-domain noise.
When in-domain noise is available, our UniVPM directly outperforms previous state-of-the-art, which breaks out the limit of data augmentation and moves one step forward to the real noisy data training setting (\textit{i.e.}, oracle).
In addition, Table~\ref{table3} further investigates the cases with out-of-domain training noise, where our UniVPM even surpasses previous state-of-the-art trained on in-domain noise.
As a result, our proposed approach effectively alleviates the limitation of in-domain noisy data in audio modality-based robust AVSR.

\subsection{Analysis of UniVPM from Meta-Learning Perspective}
\label{assec:analysis_univpm_meta}
The main idea of our proposed UniVPM can also be explained from meta-learning perspective~\citep{raghu2019rapid}, \textit{i.e.}, learn how to learn.
In AVSR task, considering the inherent information sufficiency of visual modality to represent speech content~\citep{sataloff1992human,ren2021learning}, the key factor of its robustness is still the informative audio modality.
However, audio is usually interfered by background noise during practical inference.
Therefore, the key of improving robustness is to gain sufficient knowledge from clean audio in training stage, and meta-learning exactly tells AVSR how to learn from the clean audio.
Motivated by this idea, we leverage clean audio-visual data to train the core modules of UniVPM, \textit{i.e.}, viseme and phoneme banks, where video serves as ``prompt'' and clean audio serves as ``meta''.
In particular, our UniVPM learns the mapping between visemes and phonemes, which then enables modality transfer to restore clean audio against testing noises.
Here the viseme-phoneme mapping defines \emph{how to learn from clean audio}.
Therefore, we only need video ``prompt'' during inference to access the clean audio ``meta'', which enables UniVPM to adapt to any testing noises.

\subsection{Analysis of Phoneme Distribution in LRS3 and LRS2 Datasets}
\label{assec:analysis_phoneme_distribution}

\begin{figure}[t!]
\centering
    \includegraphics[width=1.0\columnwidth]{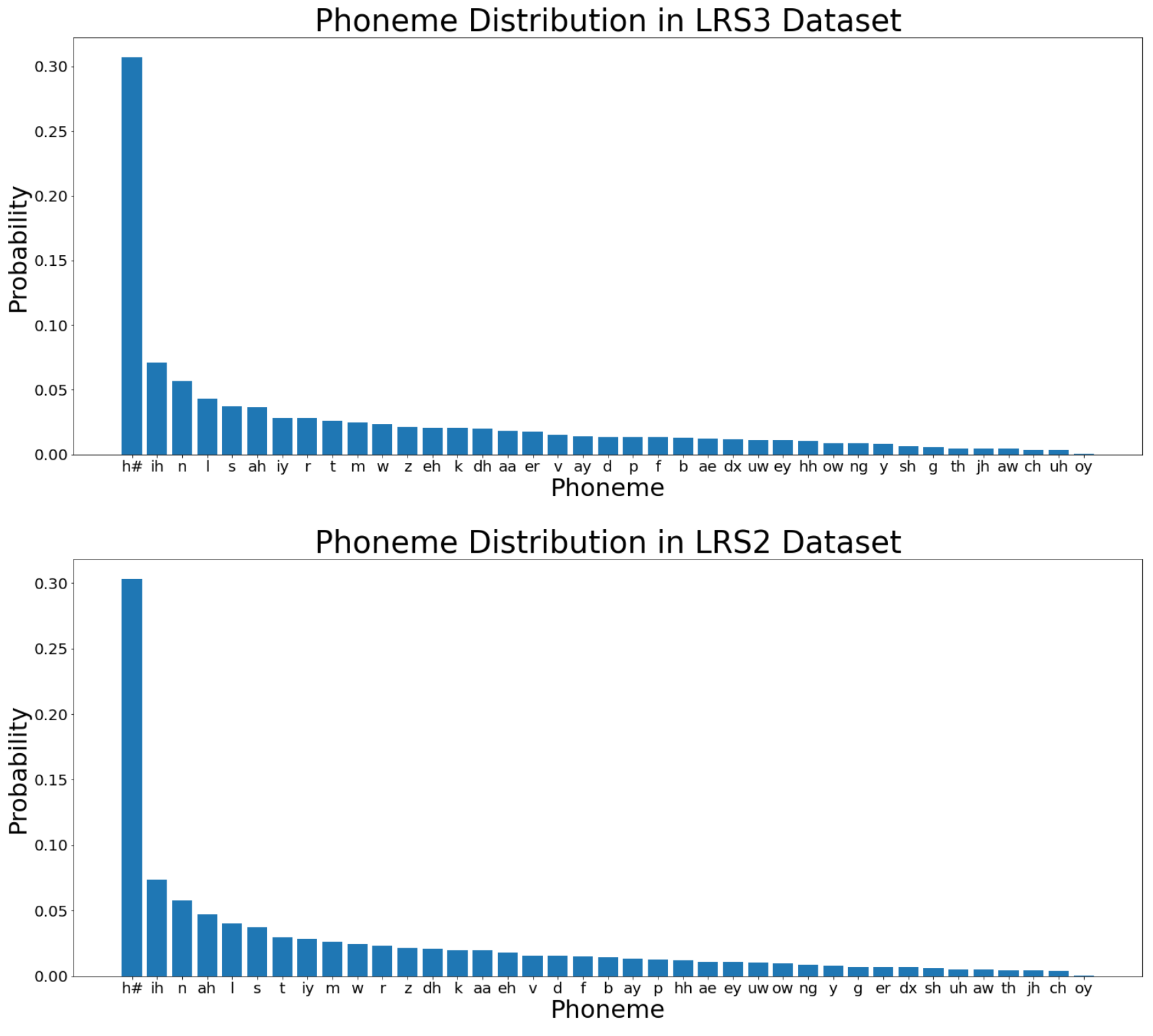}
    \vspace{-0.6cm}
    \caption{Phoneme distributions in LRS3 and LRS2 datasets. Pre-trained phoneme recognition model~\citep{phy22phoneme} is used for statistics, where speech is recognized into 44 phonemes, with 39 of them visualized in figures and another 5 special phonemes eliminated (\textit{i.e.}, `\texttt{|}', `\texttt{[UNK]}', `\texttt{[PAD]}', `\texttt{<s>}', `\texttt{</s>}').}\label{fig6}
    \vspace{-0.5cm}
\end{figure}

\begin{figure}[t!]
\centering
    \includegraphics[width=1.0\columnwidth]{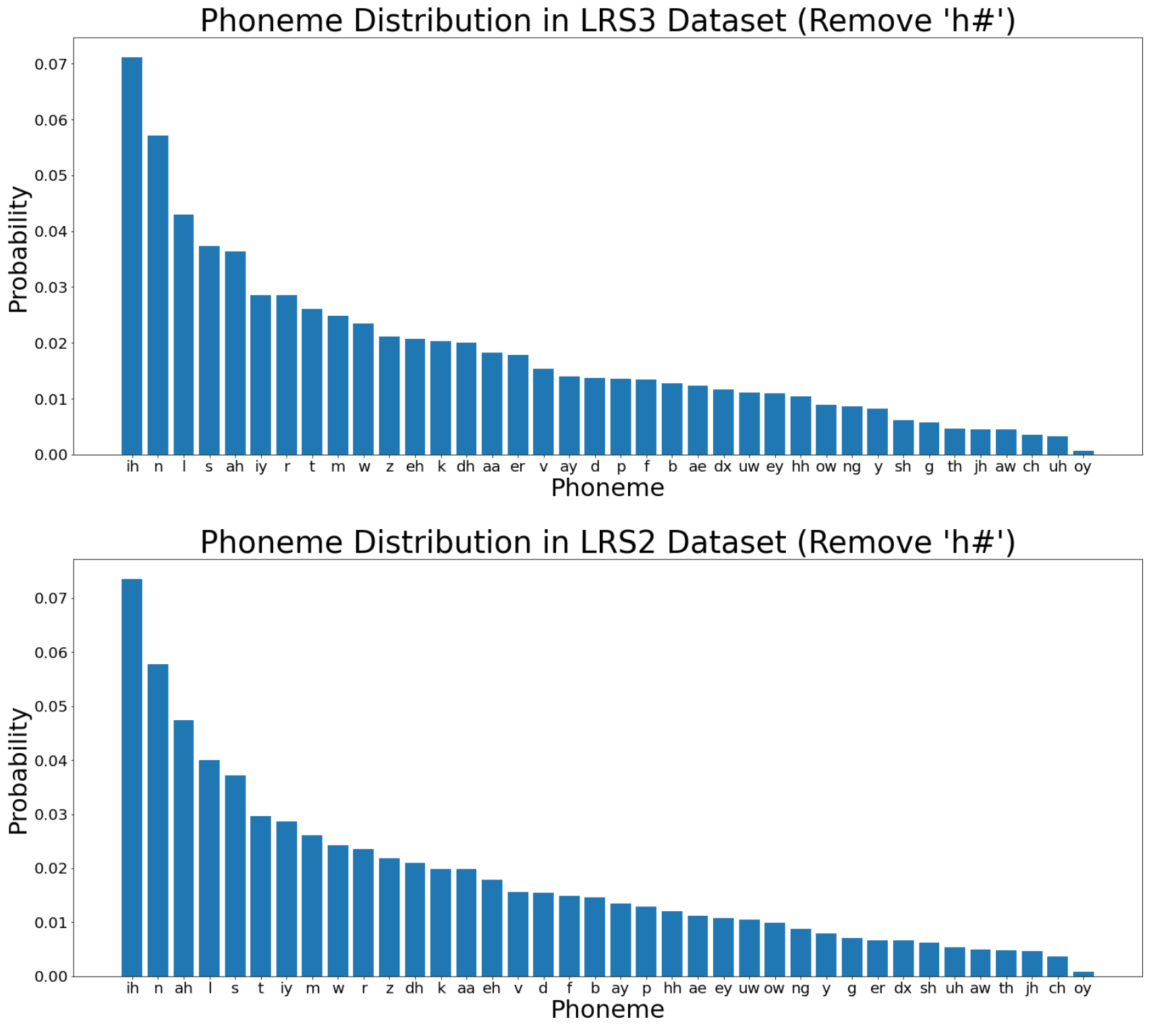}
    \vspace{-0.6cm}
    \caption{Phoneme distributions without `\texttt{h\#}'.}\label{fig7}
    \vspace{-0.1cm}
\end{figure}

Fig.~\ref{fig6} presents the phoneme distribution in LRS3 and LRS2 datasets.
We can observe that in both datasets, the phoneme obeys a long-tail distribution~\citep{liu2019large} with head classes including `\texttt{h\#}', `\texttt{ih}', `\texttt{n}', `\texttt{l}', `\texttt{s}', `\texttt{ah}', etc.
For better visualization, Fig.~\ref{fig7} removes the dominant phoneme `\texttt{h\#}' and also presents a long-tail distribution.
Therefore, the neural network trained on these data is prone to over-fitting to head phoneme classes, resulting in less satisfactory performance on tail classes.

LRS3 and LRS2 are both large-scale English reading speech datasets recorded with thousands of speakers from a wide range of races, so that they can be roughly representative of the phoneme distribution of English language.

\begin{figure}[t]
\centering
    \includegraphics[width=1.0\columnwidth]{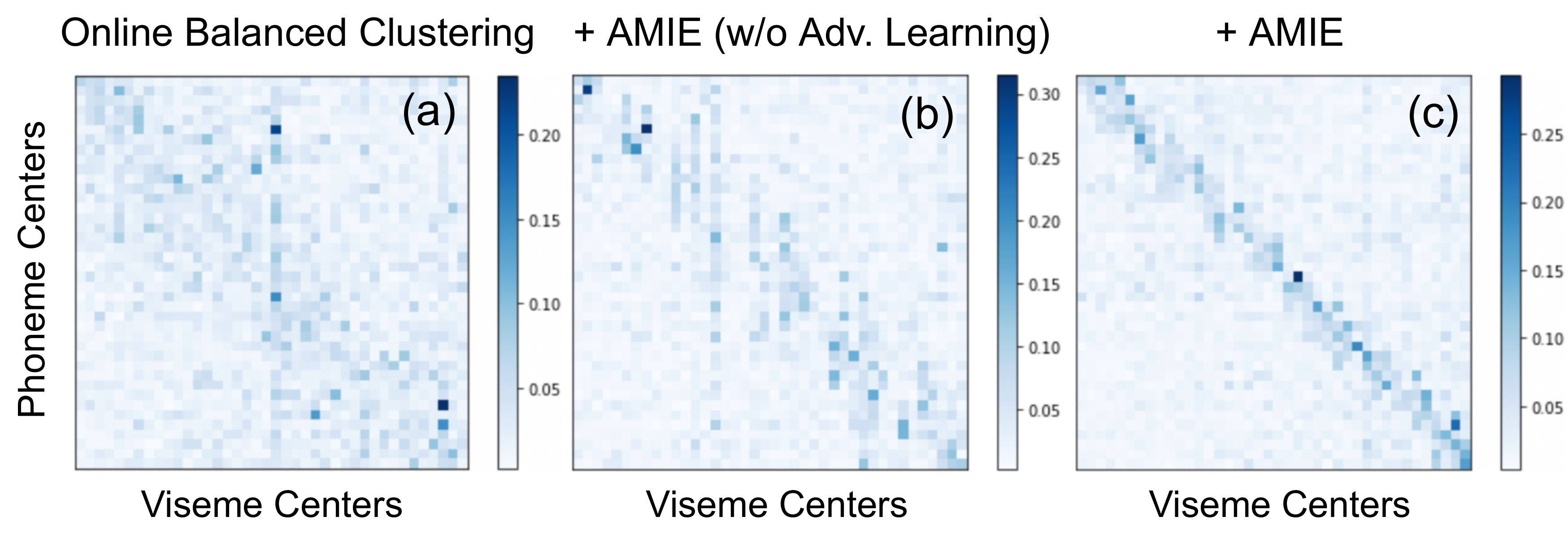}
    \vspace{-0.6cm}
    \caption{Confusion matrix of viseme-phoneme mapping in (a) Online Balanced Clustering, (b) Online Balanced Clustering + AMIE (without adversarial learning) and (c) Online Balanced Clustering + AMIE.}\label{fig8}
    \vspace{-0.3cm}
\end{figure}

\section{Experimental Details}
\label{asec:exp_details}

\subsection{Datasets}
\label{assec:datasets}
\noindent\textbf{LRS3}\footnote{\url{https://www.robots.ox.ac.uk/~vgg/data/lip_reading/lrs3.html}}~\citep{afouras2018lrs3} is currently the largest public sentence-level lip reading dataset, which contains over 400 hours of English video extracted from TED and TEDx talks on YouTube.
The training data is divided into two parts: pretrain (403 hours) and trainval (30 hours), and both of them are transcribed at sentence level.
The pretrain part differs from trainval in that the duration of its video clips are at a much wider range. 
Since there is no official development set provided, we randomly select 1,200 samples from trainval as validation set ($\sim$ 1 hour) for early stopping and hyper-parameter tuning.
In addition, it provides a standard test set (0.9 hours) for evaluation.

\noindent\textbf{LRS2}\footnote{\url{https://www.robots.ox.ac.uk/~vgg/data/lip_reading/lrs2.html}}~\citep{chung2017lip} is a large-scale publicly available labeled audio-visual (A-V) datasets, which consists of 224 hours of video clips from BBC programs.
The training data is divided into three parts: pretrain (195 hours), train (28 hours) and val (0.6 hours), which are all transcribed at sentence level.
An official test set (0.5 hours) is provided for evaluation use.
The dataset is very challenging as there are large variations in head pose, lighting conditions and genres.

\subsection{Data Preprocessing}
\label{subsec:preprocessing}
The data preprocessing for above two datasets follows the LRS3 preprocessing steps in prior work~\citep{shi2022learning}.
For the audio stream, we extract the 26-dimensional log filter-bank feature at a stride of 10 ms from input raw waveform.
For the video clips, we detect the 68 facial keypoints using dlib toolkit~\citep{king2009dlib} and align the image frame to a reference face frame via affine transformation.
Then, we convert the image frame to gray-scale and crop a 96$\times$96 region-of-interest (ROI) centered on the detected mouth.
During training, we randomly crop a 88$\times$88 region from the whole ROI and flip it horizontally with a probability of 0.5. 
At inference time, the 88$\times$88 ROI is center cropped without horizontal flipping. 
To synchronize these two modalities, we stack each 4 neighboring acoustic frames to match the image frames that are sampled at 25Hz.

\subsection{Model Configurations}
\label{assec:model_config}
\noindent\textbf{Front-ends.} 
We adopt the modified ResNet-18 from prior work~\citep{shi2022learning} as visual front-end, where the first convolutional layer is replaced by a 3D convolutional layer with kernel size of 5$\times$7$\times$7. 
The visual feature is flattened into an 1D vector by spatial average pooling in the end.
For audio front-end, we use one linear projection layer followed by layer normalization~\citep{ba2016layer}.

\noindent\textbf{UniVPM.}
The viseme and phoneme banks contain $N=40$ clusters, following the amount of English phonemes~\citep{phy22phoneme}, \textit{i.e.}, 39 regular phonemes and one special phoneme `\texttt{[PAD]}' that indicates silence.
It is worth mentioning that the actual amount of visemes is less than phonemes due to homophene phenomenon, \textit{i.e.}, one-to-many lip-audio mapping~\citep{bear2017phoneme}, but in this work we set same number of clusters to construct a strict one-to-one viseme-phoneme mapping, as shown in Fig.~\ref{fig5} and Fig.~\ref{fig8}.
The cluster capacity $S_{max}$ in Alg.~\ref{alg1} is set to 20, and the temperature $\tau$ in Eq.~\ref{eq9} is set to 0.1.

\noindent\textbf{Speech Recognition.}
The downstream speech recognition model follows AV-HuBERT~\citep{shi2022robust} with 24 Transformer~\citep{vaswani2017attention} encoder layers and 9 decoder layers, where the embedding dimension/feed-forward dimension/attention heads in each Transformer layer are set to 1024/4096/16 respectively.
We use a dropout of $p = 0.1$ after the self-attention block within each Transformer layer, and each Transformer layer is dropped~\citep{fan2019reducing} at a rate of 0.1.

The total number of parameters in our UniVPM and AV-HuBERT baseline are 478M and 476M.

\subsection{Data Augmentation}
\label{assec:data_aug}
Following prior work~\citep{shi2022robust}, we use many noise categories for data augmentation to simulate noisy training data.
We select the noise categories of ``\texttt{babble}'', ``\texttt{music}'' and ``\texttt{natural}'' from MUSAN noise dataset~\citep{snyder2015musan}, and extract some ``\texttt{speech}'' noise samples from LRS3 dataset.
For experiments on unseen testing noises (see Table~\ref{table3}), we also select the noise categories of ``\texttt{Meeting}'', ``\texttt{Cafe}'', ``\texttt{Resto}'' and ``\texttt{Station}'' from DEMAND noise dataset~\citep{thiemann2013diverse}.
All categories are divided into training, validation and test partitions.

During training process, we randomly select one noise category and sample a noise clip from its training partition. 
Then, we randomly mix the sampled noise with input clean audio, at signal-to-noise ratio (SNR) of 0dB with a probability of 0.25.

At inference time, we evaluate our model on clean and noisy test sets respectively.
Specifically, the system performance on each noise type is evaluated separately, where the testing noise clips are added at five different SNR levels: $\{-10, -5, 0, 5, 10\}dB$.
At last, the testing results on different noise types and SNR levels will be averaged to obtain the final noisy WER result.

\subsection{Training and Inference}
\label{assec:train_infer}
\noindent\textbf{Training.}
The noisy training data is synthesized by adding random noise from MUSAN~\citep{snyder2015musan} or DEMAND~\citep{thiemann2013diverse} of 0dB at a probability of 0.25.
We load the pre-trained AV-HuBERT\footnote{\url{https://github.com/facebookresearch/av_hubert}} for front-ends and downstream speech recognition model, and then follow its sequence-to-sequence (S2S) finetuning configurations~\citep{shi2022robust} to train our system.
We use Transformer decoder to decode the encoded features into unigram-based subword units~\citep{kudo2018subword}, where the vocabulary size is set to 1000.
The weighting parameters $\lambda_{G\hspace{-0.02cm}A\hspace{-0.02cm}N}/\lambda_{rec}/\lambda_{var}$ in Eq.~\ref{eq12} are set to 0.1/0.2/0.5, respectively.
The entire system is trained for 60K steps using Adam optimizer~\citep{kingma2014adam}, where the learning rate is warmed up to a peak of 0.001 for the first 20K updates and then linearly decayed.
The training process takes $\sim$ 2.5 days on 4 NVIDIA-V100-32GB GPUs, where in comparison the AV-HuBERT finetuning takes $\sim$ 1.3 days on 4 NVIDIA-V100-32GB GPUs.

\noindent\textbf{Inference.}
As shown in Table~\ref{table1}, the testing noises ``\texttt{Babble}'', ``\texttt{Music}'' and ``\texttt{Natural}'' are sampled from MUSAN, and ``\texttt{Speech}'' is drawn from LRS3, following prior work~\citep{shi2022robust}.
No language model is used during inference.
We employ beam search for decoding, where the beam width and length penalty are set to 50 and 1 respectively.
All hyper-parameters in our systems are tuned on validation set.
Since our experimental results are quite stable, a single run is performed for each reported result.

\subsection{Details of UniVPM Optimization}
\label{assec:detail_univpm_optim}

As detailed in Alg.~\ref{alg2}, we design a two-step adversarial learning strategy for UniVPM optimization, where the discriminator and generator play a two-player minimax game.
First, we \emph{maximize} $\mathcal{L}_{G\hspace{-0.02cm}A\hspace{-0.02cm}N}$ to update the discriminator, where generator is detached from optimization.
According to Eq.~\ref{eq11}, maximizing the first term of $\mathcal{L}_{G\hspace{-0.02cm}A\hspace{-0.02cm}N}$ increases the MI between visual and audio sequences, while maximizing the second term is actually decreasing the MI between visemes and phonemes, as well as the MI between visual and restored audio sequences (this is opposite to our desired viseme-phoneme mapping and modality transfer).
Second, we freeze discriminator and update the rest network, where \emph{minimizing} $\mathcal{L}_G$ increases the MI between visemes and phonemes, as well as MI between visual and restored audio sequences.
In addition, $\mathcal{L}_{A\hspace{-0.01cm}S\hspace{-0.01cm}R}$ optimizes the downstream speech recognition model, $\mathcal{L}_{rec}$ supervise the quality of restored clean audio, and $\mathcal{L}_{var}$ disperses the viseme and phoneme centers to ease their mapping construction.
The entire system is trained in an end-to-end manner.

In actual experiments, to save computation cost, we update $\mathcal{B}_v$ and $\mathcal{B}_a$ once every 10 epochs, which has been proved with no affect on the system performance.
One can refer to our attached code for more implementation details.

\begin{algorithm}[t]
\caption{\small UniVPM Optimization.}
\label{alg2}
\small 
\begin{algorithmic}[1]
    \Require Training data $D_\text{train}$ that contains visual-audio pairs $(x_v, x_a)$ and the text transcription $y$. 
    The UniVPM network $\theta$ that consists of visual front-end $\theta_{v\hspace{-0.02cm}f}$, audio front-end $\theta_{a\hspace{-0.02cm}f}$, viseme bank $\mathcal{B}_v$, phoneme bank $\mathcal{B}_a$, AMIE $\theta_{A\hspace{-0.03cm}M\hspace{-0.03cm}I\hspace{-0.03cm}E}$ and speech recognition model $\theta_{A\hspace{-0.01cm}S\hspace{-0.01cm}R}$.
    Hyper-parameter weights $\lambda_{G\hspace{-0.02cm}A\hspace{-0.02cm}N}$, $\lambda_{rec}$, $\lambda_{var}$.
    \State Load pre-trained AV-HuBERT for $\theta_{v\hspace{-0.02cm}f}$, $\theta_{a\hspace{-0.02cm}f}$ and $\theta_{A\hspace{-0.01cm}S\hspace{-0.01cm}R}$, randomly initialize $\theta_{A\hspace{-0.03cm}M\hspace{-0.03cm}I\hspace{-0.03cm}E}$.
    \State Initialize empty banks $\mathcal{B}_v$ and $\mathcal{B}_a$.
    \While{not converged}
        \For{$(x_v, x_a) \in D_\text{train}$}
            \State \textsc{Forward-Propagation}:
            \State \hspace{0.3cm} $f_v = \theta_{v\hspace{-0.02cm}f}(x_v), f_a = \theta_{a\hspace{-0.02cm}f}(x_a)$ \Comment{\blue{front-ends}}
            \State \hspace{0.3cm} Update $\mathcal{B}_v$ and $\mathcal{B}_a$ according to Alg.~\ref{alg1}
            \State \hspace{0.3cm} Obtain viseme sequence $s_v$ from $f_v$ and $\mathcal{B}_v$
            \State \hspace{0.3cm} Obtain phoneme sequence $s_a$ from $f_a$ and $\mathcal{B}_a$
            \State \hspace{0.3cm} Generate restored audio $\hat{f}_a$ in Eq.~\ref{eq9} and~\ref{eq10}
	        \State \hspace{0.3cm} $\hat{y} = \theta_{A\hspace{-0.01cm}S\hspace{-0.01cm}R}(f_v\oplus f_a\oplus \hat{f}_{a})$ \Comment{\blue{recognition}}
	        \State \textsc{Training objectives}:
	        \State \hspace{0.3cm} $\mathcal{L}_{G\hspace{-0.02cm}A\hspace{-0.02cm}N}$ ($\mathcal{L}_D$ and $\mathcal{L}_G$) in Eq.~\ref{eq11}
	        \State \hspace{0.3cm} $\mathcal{L}_{rec} = \Vert \hat{f}_a - f_a\Vert_2$
	        \State \hspace{0.3cm} $\mathcal{L}_{var} = \text{Var}(c_v^1, ..., c_v^N) + \text{Var}(c_a^1, ..., c_a^N)$
	        \State \hspace{0.3cm} $\mathcal{L}_{A\hspace{-0.01cm}S\hspace{-0.01cm}R} = \text{CrossEntropy}(\hat{y}, y)$
	        \State \textsc{Back-Propagation}: \Comment{\blue{adversarial learning}}
	        \State \hspace{0.3cm} \textsc{Update AMIE}: \Comment{\blue{unfreeze $\theta_{A\hspace{-0.03cm}M\hspace{-0.03cm}I\hspace{-0.03cm}E}$}}
	        \State \hspace{0.6cm} $\underset{\theta_{A\hspace{-0.03cm}M\hspace{-0.03cm}I\hspace{-0.03cm}E}}{\arg\max}\hspace{0.1cm} \mathcal{L}_{G\hspace{-0.02cm}A\hspace{-0.02cm}N}$ 
	        \State \hspace{0.3cm} \textsc{Update rest network}: \Comment{\blue{freeze $\theta_{A\hspace{-0.03cm}M\hspace{-0.03cm}I\hspace{-0.03cm}E}$}}
	        \State \hspace{0.6cm} $\underset{\theta\backslash\theta_{A\hspace{-0.03cm}M\hspace{-0.03cm}I\hspace{-0.03cm}E}}{\arg\min}\hspace{0.1cm} \mathcal{L}_{A\hspace{-0.01cm}S\hspace{-0.01cm}R} + \lambda_{G\hspace{-0.02cm}A\hspace{-0.02cm}N} \cdot \mathcal{L}_G + \lambda_{rec} \cdot \mathcal{L}_{rec} + \lambda_{var} \cdot \mathcal{L}_{var}$ 
        \EndFor
    \EndWhile
\end{algorithmic}
\normalsize
\end{algorithm}

\subsection{Baselines}
\label{assec:baselines}
In this section, we describe the baselines for comparison.

\begin{itemize}
\item \textbf{TM-seq2seq}~\citep{afouras2018deep}: TM-seq2seq proposes a Transformer-based AVSR system to model the A-V features separately and then attentively fuse them for decoding, and uses cross-entropy based sequence-to-sequence loss as training criterion.
\item \textbf{TM-CTC}~\citep{afouras2018deep}: TM-CTC shares the same architecture with TM-seq2seq, but uses CTC loss~\citep{graves2006connectionist} as training criterion.
\item \textbf{Hyb-RNN}~\citep{petridis2018audio}: Hyb-RNN proposes a RNN-based AVSR model with hybrid seq2seq/CTC loss~\citep{watanabe2017hybrid}, where the A-V features are encoded separately and then concatenated for decoding.
\item \textbf{RNN-T}~\citep{makino2019recurrent}: RNN-T adopts the popular recurrent neural network transducer~\citep{graves2012sequence} for AVSR task, where the audio and visual features are concatenated before fed into the encoder.
\item \textbf{EG-seq2seq}~\citep{xu2020discriminative}: EG-seq2seq builds a joint audio enhancement and multimodal speech recognition system based on RNN~\citep{zhang2019eleatt}, where the A-V features are concatenated before decoding.
\item \textbf{LF-MMI TDNN}~\citep{yu2020audio}: LF-MMI TDNN proposes a joint audio-visual speech separation and recognition system based on time-delay neural network (TDNN), where the A-V features are concatenated before fed into the recognition network.
\item \textbf{Hyb-AVSR}~\citep{ma2021end}: Hyb-AVSR proposes a Conformer-based~\citep{gulati2020conformer} AVSR system with hybrid seq2seq/CTC loss, where the A-V input streams are first encoded separately and then concatenated for decoding.
\item \textbf{MoCo+w2v2}~\citep{pan2022leveraging}: MoCo+w2v2 employs self-supervised pre-trained audio and visual front-ends, \textit{i.e.}, wav2vec 2.0~\citep{baevski2020wav2vec} and MoCo v2~\citep{chen2020improved}, to generate better audio-visual features for fusion and decoding.
\item \textbf{AV-HuBERT}~\citep{shi2022learning, shi2022robust}: AV-HuBERT employs self-supervised learning to capture deep A-V contextual information, where the A-V features are masked and concatenated before fed into Transformer encoder to calculate masked-prediction loss for pre-training, and sequence-to-sequence loss is then used for finetuning.
\item \textbf{u-HuBERT}~\citep{hsu2022u}: u-HuBERT extends AV-HuBERT to a unified framework of audio-visual and audio-only pre-training.
\item \textbf{Distill-PT}~\citep{ma2022visual}: Distill-PT proposes a Conformer-based VSR framework with additional distillation from pre-trained ASR and VSR models.

\end{itemize}

\section{Clustering Algorithms}
\label{asec:cluster_alg}

\subsection{Uniform Effect in $K$-Means}
\label{assec:uniforme_effect}
$K$-Means~\citep{macqueen1967classification} is the most popular and successful clustering algorithm, where sample re-allocation and center renewal are executed alternatively to minimize the intra-cluster distance.
However,~\citet{xiong2006k} points out that $K$-Means algorithm tends to produce balanced clustering result, \textit{a.k.a.}, uniform effect.
This preference seriously degrades the performance when the clusters are imbalanced-sized. 
The consequence is that the center of minority clusters will gradually move to the territory of majority cluster, as illustrated in Fig.~\ref{fig3} (a).
In other words, the $K$-Means algorithm will be over-fitted to majority clusters, leaving the samples in minority clusters not well modeled.

\subsection{$K$-Means++}
\label{assec:kmeans++}
The performance of $K$-Means clustering relies on the center initialization, where the vanilla algorithm initialize cluster centers randomly.
$K$-Means++~\citep{arthur2006k} is an improved version with dispersed initial centers.
It determines cluster centers one by one, and each newly initialized center is pushed as distant as possible to the existed centers.
As a result, the $K$ initial cluster centers would separate from each other and benefit the subsequent clustering process.

\subsection{Details of Online Clustering Baseline}
\label{assec:detail_online_cluster}
For comparison, we build an Online Clustering algorithm as baseline.
It is similar to Alg.~\ref{alg1} but employs a vanilla random pruning strategy, instead of re-sampling, to control the total memory of the bank.
Our strategy is to randomly keep $S_{thr}$ samples in the cluster if its number of samples exceeds $S_{thr}$.
Compared to the proposed Online Balanced Clustering algorithm, this baseline also controls memory size but ignores the imbalanced clusters, as indicated by the dashed ellipses in Fig.~\ref{fig3} (a).

\subsection{Principles of Online Balanced Clustering}
\label{assec:principle_obc}
According to Alg.~\ref{alg1}, the main idea of proposed Online Balanced Clustering is the re-sampling operation to balance cluster sizes.
For majority clusters, we perform undersampling to maintain the $S_{thr}$ nearest samples to cluster center, so that the gathered clusters in Fig.~\ref{fig3} (a) can be separated.
For minority clusters, we introduce oversampling to interpolate a new sample near the center, so that the minority clusters are highlighted.
As a result, all the clusters are balanced-sized and separated from each other as shown in Fig.~\ref{fig3} (b), so that the over-fitting problem is resolved.
As a result, all of the visemes and phonemes can get well represented, which enables the subsequent viseme-phoneme mapping construction.

\begin{figure*}[t]
\centering
    \includegraphics[width=0.92\textwidth]{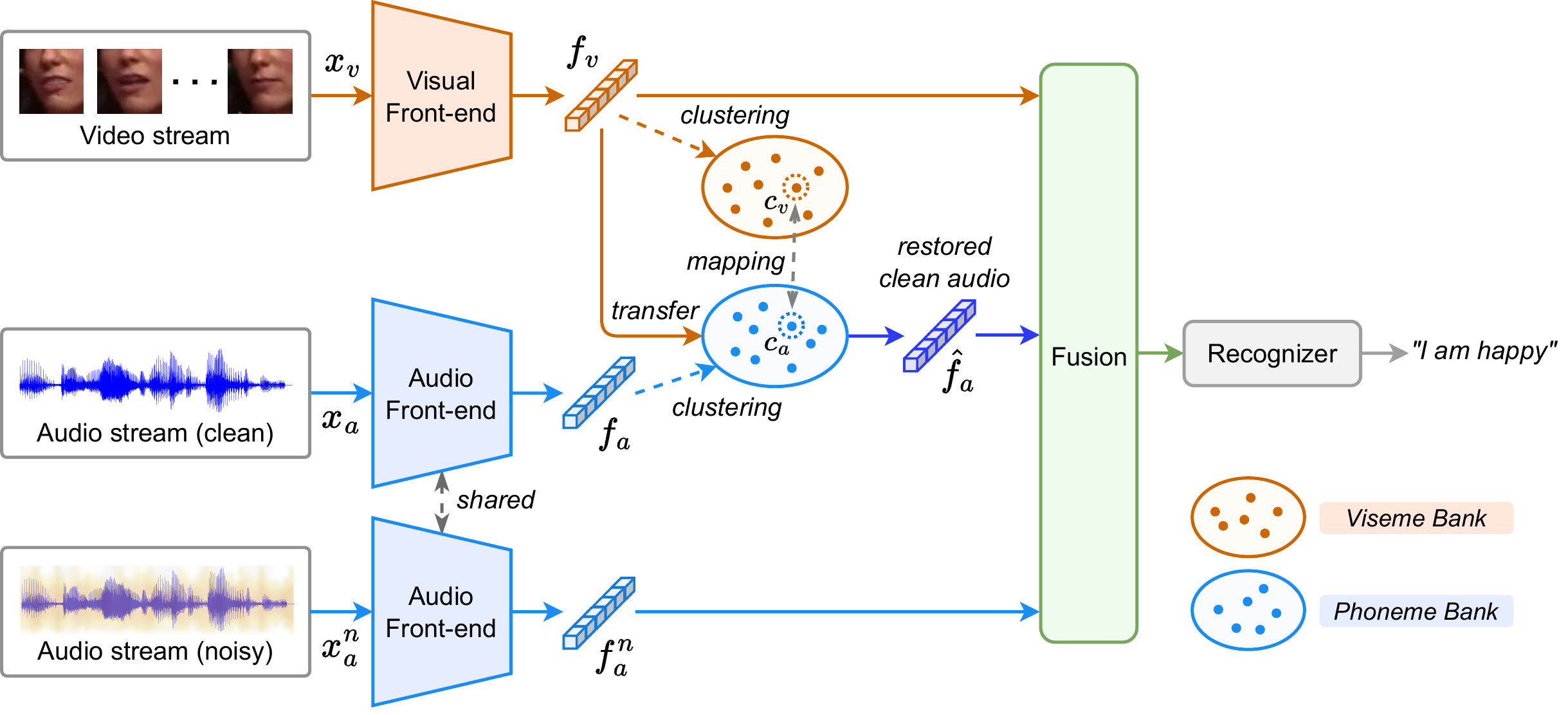}
    \vspace{-0.1cm}
    \caption{Illustration of noisy training pipeline of UniVPM. 
    Both clean and noisy audio are used for training, where the clean audio is employed for phoneme clustering and the noisy audio is used to improve the system noise-robustness.
    Compared to Fig.~\ref{fig2}, there is an extra data stream of noisy audio to improve robustness.
    }\label{fig9}
    \vspace{-0.2cm}
\end{figure*}

\end{document}